\documentclass[american,doc,11pt,floatsintext]{apa7}
\usepackage{mathptmx}
\usepackage{expex}
\usepackage[T1]{fontenc}
\usepackage{longtable}
\usepackage{booktabs}
\usepackage{mathtools}
\usepackage{amsmath}
\usepackage{graphicx}
\PassOptionsToPackage{normalem}{ulem}
\usepackage{ulem}
\usepackage[utf8]{inputenc}
\usepackage{babel}
\usepackage{csquotes}
\usepackage{tikz-dependency}
\usepackage[style=apa,natbib=true]{biblatex}
\hypersetup{breaklinks=false,hidelinks}

\newcommand{\mr}[1]{#1}
\newcommand{\mrtwo}[1]{#1}

\begin{document}
\title{SEAM: An Integrated Activation-Coupled Model of Sentence Processing and Eye Movements in Reading}
\shorttitle{Sentence Processing and Eye Movements}
\authorsnames[1,2,{2,4},{2,3},{1,3}]{Maximilian~M.~Rabe, Dario~Paape, Daniela~Mertzen, Shravan~Vasishth, Ralf~Engbert}
\leftheader{Rabe et al.}
\authorsaffiliations{{Department of Psychology, University of Potsdam},{Department of Linguistics, University of Potsdam},{Research Focus Cognitive Sciences, University of Potsdam},{University Library, University of Potsdam}}
\widowpenalty10000
\authornote{\addORCIDlink{Maximilian M. Rabe}{0000-0002-2556-5644}\par
\addORCIDlink{Dario Paape}{0000-0002-7148-5258}\par
\addORCIDlink{Daniela Mertzen}{0000-0003-4471-9255}\par
\addORCIDlink{Shravan Vasishth}{0000-0003-2027-1994}\par
\addORCIDlink{Ralf Engbert}{0000-0002-2909-5811}\par
\textbf{Corresponding Author.} Correspondence should be addressed to Maximilian M.\ Rabe, Universität Potsdam, Karl-Liebknecht-Str.\ 24--25, 14476 Potsdam, Germany, \href{mailto:maximilian.rabe@uni-potsdam.de}{maximilian.rabe@uni-potsdam.de}. \par
\textbf{Data Availability.} All data, analysis code, and computational models used in this manuscript are available at the Open Science Framework \mrtwo{(\url{https://doi.org/10.17605/OSF.IO/8ZRXB})} and at \mrtwo{the University of Potsdam (\url{https://gitup.uni-potsdam.de/CRC1294/B03/SEAM-2023-Paper})}.
}


\abstract{Models of eye-movement control during reading, developed largely within psychology, usually focus on visual, attentional, \mr{\label{r1-1.1}lexical,} and motor processes but neglect post-lexical language processing; by contrast, models of sentence comprehension processes, developed largely within psycholinguistics, generally focus only on post-lexical language processes. We present a model that combines these two research threads, by integrating eye-movement control and sentence processing. Developing such an integrated model is extremely challenging and computationally demanding, but such an integration is an important step toward complete mathematical models of natural language comprehension in reading. We combine the SWIFT model of eye-movement control \mr{(\citeauthor{Seelig2020}, \emph{Journal of Mathematical Psychology, 95}, \citeyear{Seelig2020}, Article 102313)} with key components of the Lewis and Vasishth sentence processing model (\citeauthor{Lewis2005}, \emph{Cognitive Science, 29}, \citeyear{Lewis2005}, pp.~375--419). This integration becomes possible, for the first time, due in part to recent advances in successful parameter identification in dynamical models, which allows us to investigate profile log-likelihoods for individual model parameters. We present a fully implemented proof-of-concept model demonstrating how such an integrated model can be achieved; our approach includes Bayesian model inference with Markov Chain Monte Carlo (MCMC) sampling as a key computational tool. The integrated \mrtwo{Sentence-Processing and Eye-Movement Activation-Coupled Model} (SEAM) can successfully reproduce eye movement patterns that arise due to similarity-based interference in reading. To our knowledge, this is the first-ever integration of a complete process model of eye-movement control with linguistic dependency completion processes in sentence comprehension. In future work, this proof of concept model will need to be evaluated using a comprehensive set of benchmark data.}
\maketitle

\section{Introduction}

What is the relationship between sentence processing and eye movements during reading? As an answer to this question, \citet[pp.~330--331]{Just1980} famously coined the eye-mind assumption, which states that ``the eye remains fixated on a word as long as the word is being processed'', and that ``there is no appreciable lag between what is being fixated and what is being processed''. But what does it mean for a word to be ``processed''? \citeauthor{Just1980}'s model of reading has three stages: Encoding of the word form and lexical access, identification of relationships between the words in a sentence (such as agent-action-object), and integration with information from previous sentences. Once these three stages are finished, the eyes proceed to the next word.\footnote{There is a fourth stage in the  model, called wrap-up, which only occurs at the end of a sentence, and whose purpose is to finish any processing that could not be completed at a previous point during reading \citep[but see][for a critical discussion]{Warren2009}.} \citeauthor{Just1980}'s processing model is highly serial, which matches most readers' subjective experience that sentences are processed in an incremental, left-to-right fashion \citep{Snell19}. However, while readers do tend to make fixations incrementally in the reading direction, fixation sequences are not always in serial order: Instead of systematically shifting the gaze from one word to the next -- something that only happens in about 50\% of fixations -- readers also skip words, refixate the same word, or regress to previous words \mr{\citep{Kliegl2004,Rayner1998}}. 

This more complicated picture of reading aligns with the fact that the structure of many sentences in natural language does not correspond to simple agent-action-object sequences. Consider a sentence like (\ref{ex1}), taken from \citet{Mertz21}:

\ex\label{ex1} It turned out that the attorney whose secretary had forgotten that the visitor was important frequently complained about the salary at the firm.\xe

In this sentence, there are several dependencies between non-adjacent words, most strikingly the long-distance dependency between the noun \textit{attorney} and the verb \textit{complained}. It is difficult to argue that the processing of the word \textit{attorney} is finished once the preamble \textit{It turned out that the attorney~\ldots} has been read: It is clear that a verb must arrive at some point of which \textit{attorney} is the subject. Complete integration of \textit{attorney} can thus only be achieved when \textit{complained} is read after ten intervening words have been processed. It is therefore clear that the eyes will have to move forward even if the current word has not been completely integrated into the sentence structure. 

A well-established assumption in sentence processing is that a noun like \textit{attorney} is held in working memory until the dependency is completed, and needs to be retrieved when the verb is reached \citep{gibson98,gibson00,Lewis06}. A strong interpretation of the eye-mind assumption would predict that, given that the processing of \textit{attorney} is finalized at \textit{complained}, readers should refixate \textit{attorney} once lexical access of \textit{complained} is complete. However, this is not what usually happens: While readers do make more regressions in more complex sentences that involve memory retrievals (e.g., \citealp{Gordon2006,Lee07,Mertz21,Jaeg15}), regressive eye movements nevertheless occur only in a minority of trials. \mr{\label{r1-0.2b}Furthermore, even in difficult sentences that may require multiple passes to parse correctly, readers do not necessarily regress to the most syntactically informative words in the sentence} (e.g, \citealp{malsburgvasishthsub10,MalsburgVasishth2013,Engelmann2013,Chris17}). Thus, while there is undoubtedly a connection between sentence processing and eye movements \citep{Rayner1998,cliftonjr:emr,Frazier1982}, it is much less direct than posited by the strong version of the eye-mind assumption, as \citet{Reichle09} have pointed out. \mr{On the other hand, there \textit{is} evidence that readers can and do move their eyes into the vicinity of critical words \citep{Mitchell08,Inhoff2005,Weger2007,Schotter14,Mese02}, which suggests the need for a model with \textit{some} linguistically-mediated guidance of regressive eye movements.}

Psycholinguistic studies of sentence processing typically rely on aggregated reading measures such as total fixation times, and models of language processing during reading, such as the classic \citet{Just1980} model, usually ignore the complexity of eye-movement control. However, highly detailed models of eye-movement control do exist. An important line of work in cognitive psychology seeks to explain reading processes at the level of individual fixations and saccades by unpacking the underlying dynamics of the latent sub-processes involved. Several influential mathematical models of eye-movement control exist; a prominent example is the E-Z Reader model \citep{Reichle2003}. These models have historically focused on the effects of word-level properties such as word length, frequency, and predictability, and do not take into account higher-level processes such as linguistic dependency completion. However, there have been several attempts at integrating models of sentence processing difficulty with eye-movement control, including E-Z Reader \citep{Reichle09}, the model of \citet{Engelmann2013}, and Über-Reader \citep{Reichle2020, Veldre2020}. These models focus on different aspects of sentence processing, and have been evaluated against corpus data, such as the Schilling corpus \citep{schilling1998cnl}. Two models that investigate the interaction between eye-movement control and sentence comprehension using data from planned experiments are reported in \citet{VasishthEngelmann2020} and \citet{Dotlacil2021}; both these investigations use a highly simplified version of E-Z Reader, that is, the Eye Movements and Movement of Attention (EMMA) model embedded within the ACT-R architecture \citep{Salvucci2001}\label{r1-0.1-emma}. The simplified EMMA model has important limitations; for example, as discussed in \citet{Engelmann2013}, the model only allows regressive eye movements to the preceding word.

All of these existing models do capture a range of selected empirical phenomena and furnish important insights into the interaction between eye-movement control and sentence parsing processes. However, to our knowledge, no model exists that uses a fully specified mainstream model of eye-movement control that is integrated with a model of dependency completion in language comprehension; furthermore, as far as we are aware, such a detailed process model has never been evaluated using data from a planned psycholinguistic experiment.

A major difficulty in developing a more complex integrated model is that a considerable number of  model parameters will need to be estimated using empirical data. For models of such complexity, conventional methods like grid search will lead to intractability. In order to implement such a complex model, Bayesian parameter estimation using the model's likelihood function (or an approximation) provides a rigorous approach to statistical inference \citep{Rabe2021,Schuett2017}. 
Two major advantages of the Bayesian approach are that parameters can be regularized or constrained a priori, which makes computation more efficient compared to the traditional grid search method, and that the uncertainty of the parameter estimates can be taken into account when evaluating model fit. Regularization makes parameter estimation more tractable, and incorporating the uncertainty of parameter estimates gives a more realistic picture of model fit \citep{NicenboimEtAlBayes2019}. Although Bayesian model fitting has been implemented for a basic reading model \citep{Dotlacil2018}, this line of work currently still neglects many low-level physiological and higher-level cognitive aspects of reading. 

In this context, the major recent advance in Bayesian parameter inference for modeling process-based models has been proposed by \citet{Seelig2020,Rabe2021} \mr{\citep[for an overview, see][]{Engbert2022}}. This line of work relies on the dynamical model of eye movement control developed by \citet{Engbert2005}, and demonstrates how the Bayesian approach can be deployed in highly complex process models. \mr{\label{r1-0.1}Compared to other models of eye-movement control in reading such as E-Z Reader \citep{Reichle2003}, the Saccade-Generation With Inhibition by Foveal Targets (SWIFT) model has several advantages that make it \mrtwo{a better suited framework to investigate higher-level processing in model of gaze control}: It (1) is available for Bayesian parameter inference due to the likelihood implementation \citep{Seelig2020,Rabe2021}, (2) has a time-dependent word-activation field that can serve as the basis for memory encodings, and (3) has mechanisms that allow for long-range regressions, which are of particular interest when investigating dependencies that span several words. SWIFT and E-Z Reader also differ with regard to theoretical assumptions such as serial vs.\ parallel processing of words, but these are not our primary focus.} \mrtwo{Advances in combining Bayesian inference with dynamical modeling by} \citet{Seelig2020,Rabe2021} \mrtwo{have enabled us to find an objective answer to the question}: Can the complex lower-level cognitive and physiological principles of eye movements be integrated with a computational model of higher-level linguistic processing, taking into account the cost of long-distance dependency completion?
 
Below, we present the Sentence-Processing and Eye-Movement Activation-Coupled Model (SEAM), a novel integrated model of sentence processing and eye movement control in reading. By combining SWIFT with the cue-based memory retrieval model proposed by \citet{Lewis2005}, we can integrate spatially-distributed processing in eye movement control with rule-based dependency completion in a Bayesian model-fitting framework. We carry out model simulation using a principled Bayesian workflow \citep{Schad2020} to demonstrate the activation-based coupling between SWIFT and the \citet{Lewis2005} model. As a result, our model yields reliable Bayesian parameter estimates by generating simulated data with known parameters, and then recovering these parameters using the Bayesian parameter estimation approach. 

We also fit SEAM to recently-published empirical data from an eye-tracking experiment investigating similarity-based interference \citep{Mertz21}, providing model-driven explanations for the observed eye movement patterns. Given that SEAM simulates time-ordered fixation sequences, the model makes predictions for all spatial and temporal summary statistics that are relevant in the reading research literature (e.g., fixation probabilities, landing positions/saccade amplitudes, and fixation durations/reading times). This capability of the SEAM architecture makes it an important candidate model for theory development in psycholinguistics.

We will first introduce the \citet{Lewis2005} model of sentence processing, then introduce the basic workings of SWIFT, and finally proceed to our integrated model SEAM.

\subsection{The Activation-Based Model of Sentence Processing (Lewis \& Vasishth, 2005)}
During sentence reading, the human sentence processor has to incrementally integrate individual words into a syntactic structure, based on which sentence meaning can be derived.  \citet{Lewis2005} proposed a model of sentence processing (hereafter, we refer to this model as LV05) that is based on the cognitive architecture ACT-R \citep{Anderson1998,Anderson2005}. In the LV05 model, incoming words are incrementally integrated into syntactic constituents that are stored in memory as \emph{chunks}. Memory chunks in LV05 carry information in the form of features, which can be used to access them in memory later on. 
Chunks also have fluctuating activation values that are determined by recency and by cue match during retrieval events. For instance, in a sentence like (\ref{exr}), as the sentence is read word-by-word, the noun phrases \textit{the robber} and \textit{the policeman} are stored as memory chunks as soon as they are read. The verbs \textit{chased} and \textit{escaped} then each trigger retrievals of their respective arguments from memory.

\ex\label{exr}
\parbox{\textwidth}{\begin{dependency}[arc edge, arc angle=60]

\begin{deptext}
The robber \& that \&  the policeman \& in the patrol car \& chased \&  escaped.\\
\end{deptext}
\depedge{5}{3}{subject}
\depedge{5}{1}{object}
\depedge{6}{1}{subject}
\end{dependency}}
\xe

Taking the retrieval at the verb \emph{escaped} as an example, the dependency needs to be completed by searching working memory for a suitable memory chunk to serve as a syntactic subject. The search process is cue-based, that is, the verb specifies a set of linguistic features such as $\pm\mathrm{noun}$ or $\pm\mathrm{animate}$ to identify the correct dependent, and existing memory chunks are reactivated based on their feature specifications. The best-matching candidate is usually retrieved, but because memory activation is noisy, misretrievals occasionally occur. In addition, processing is slowed when multiple memory chunks, such as \emph{the robber} and \emph{the policeman} in (\ref{exr}), match the retrieval cues and compete for activation, which is called the fan effect \citep[e.g.,][]{Anderson1990}.

In LV05, the latency of a given retrieval is governed by a set of equations taken from the ACT-R architecture \citep{abbl02}, which determine each chunk's activation at a given point in time. Suppose that a noun phrase, say \textit{the robber} in (\ref{exr}), has been stored in memory as memory chunk $k$. When a retrieval is triggered while processing word $n$ (\textit{escaped}) later on, chunk $k$'s activation value at word $n$ is calculated as
\begin{equation}
\label{eq:memory-activation}
A_{k,n}\left(t\right)=S_{k}\left(t\right)+P_{k}\left(t\right)+B_{k}\left(t\right)\;,
\end{equation}
\noindent
where $S_{k}$ is the memory association strength, $P_{k}$ is the mismatch penalty, and $B_{k}$ is the chunk-specific \mr{base-level activation}. The fan effects $\phi_{kl}\left(t\right)$ of competing retrieval candidates of all $l$ features of memory chunk $k$ decrease the chunk's activation strength, which also depends on the $S_{\mathrm{max}}$ (\emph{maximum activation strength}) parameter, i.e., 
\begin{equation}
\label{eq:memory-association-strength}
S_{k}\left(t\right)=\sum_{l}\left[S_{\mathrm{max}}-\log\phi_{kl}\left(t\right)\right]\;.
\end{equation}
The fan effect variable $\phi_{kl}\left(t\right)$ is defined as the number of memory chunks with feature $l$ at time $t$, including memory chunk $k$ itself so that $\phi_{kl}\left(t\right)\geq1$.

The mismatch penalty decreases activation for all retrieval cues $l$ that do not match the corresponding feature of memory chunk $k$, i.e.,
\begin{equation}
P_{k}\left(t\right)=\sum_{l}\Delta_{kl}\;,
\end{equation}
where
\begin{equation}
\Delta_{kl}\coloneqq\begin{cases}
0 & \textrm{if }\mbox{cue}_{l}=\mbox{feature}_{kl}\\
-p & \textrm{otherwise}
\end{cases} \;
\end{equation}
and $p\geq0$ is a free parameter specifying the mismatch penalty incurred by each unmatched feature.

Chunks become active when words are encoded or when retrievals are performed, and then start to decay. The resulting \mr{base-level activation} at time $t$ is given by
\begin{equation}
B_{k}\left(t\right)=\sum_{i}\textrm{exp}\left(-d\cdot t-t_{ik} \right)\;
\end{equation}
where $d$ is a decay parameter and $t_{ik}$ is the $i$-th memory access (encoding or retrieval) of memory chunk $k$. 

\mr{Note that in our implementation, in contrast to the original LV05 model, $S_k$, $\phi_{kl}$, and $P_k$ are functions of time. This is because the memory schedule, that is, the set of words encoded in memory chunks, changes dynamically each time a word is encoded in memory. As encodings can happen at any time $t$, the memory schedule, and therefore the predicted fan effects and penalties, may change even while a retrieval is ongoing. This assumption is necessary to allow for dependency resolution in the case that a retrieval trigger is processed before a potential target has been stored in memory.}

Activation values are subject to stochastic noise controlled by the \emph{ans (activation noise)} parameter, so that
\begin{equation}
A_{k,n}'\left(t\right)\sim\mathrm{Logistic}\left(A_{k,n}\left(t\right),\mathit{ans}\right)\;.
\end{equation}
The memory chunk $k_{n}^{\star}$ with the highest memory activation $A_{k,n}'$ is matched for the retrieval $n$, and the retrieval latency is computed as
\begin{equation}
t_{k,n}=F\cdot\exp\left[-A_{k,n}'\left(t\right)\right]\;,\label{eq:retrieval-latency}
\end{equation}
where $F$ is the \emph{latency factor}, a free linear scaling parameter. 

Equation \eqref{eq:retrieval-latency} can be used to make quantitative predictions for reading times, and the LV05 model has been used to model a variety of phenomena in the sentence-processing literature \citep[for a review, see][]{Engelmann2019,VasishthEngelmann2020}.  However, the LV05 model can only be straightforwardly applied to paradigms in which sentences are read strictly incrementally, such as self-paced reading: The model can create chunks, track their activations, and integrate them with each other via retrievals, but it does not account for eye fixations, and cannot capture cases in which the order of fixations mismatches the serial word order due to skippings and regressions. To fully capture ``natural'' sentence reading, the LV05 model thus needs to be interactively integrated with a model that accounts for spatial and temporal aspects of eye movements.

The dynamical SWIFT model \citep{Engbert2002,Engbert2005} is a good candidate for integration with the LV05 model. Its main advantages are that it
\begin{seriate}
\item has recently been implemented for Bayesian parameter inference \citep{Seelig2020,Rabe2021},
\item predicts and explains all empirically observable saccades in sentence reading, and
\item allows for (but does not enforce) parallel processing of words.
\end{seriate}
Even though SWIFT itself does not follow an ACT-R based architecture like EMMA \citep{Salvucci2001,Engelmann2013,VasishthEngelmann2020}, an integration with ACT-R-based models such as LV05 is possible via activation-based coupling, as we will detail below after a brief introduction of SWIFT.

\subsection{The SWIFT Model of Eye-Movement Control (Engbert et al., 2005)}
SWIFT is a model of eye-movement control in reading implemented in a dynamical cognitive modeling framework \citep{Beer2000,Engbert2021}. At its core, its internal timing processes and word activations govern the temporal control and target selection for saccadic eye movements. Words with high activation values are more likely to be selected as saccade targets. SWIFT assumes that all words that fall within a \emph{processing span} around the current fixation location are processed in parallel \citep{Engbert2002}.\footnote{Other examples of parallel processing models include \emph{Glenmore} \citep{Reilly2006} and \emph{OB1-Reader} \citep{Snell2018}. These models contrast with sequential attention shift models such as \emph{E-Z Reader} \citep{Reichle1998}.} The processing rate $\Lambda_{j}\left(t\right)$ of any given word $j$ at time $t$ depends on a number of factors such as gaze eccentricity, that is, the distance between word $j$ and the currently fixated word, such that words that are further away from the visual focus are processed more slowly.

In SWIFT, each word in the sentence passes through a \emph{lexical} and \emph{post-lexical} processing stage. During lexical processing, word recognition and identification take place. As word recognition is ongoing, the \mr{discrete} activation associated with the processed word $j$, \mr{$n_{j}\left(t\right)$}, rises up to a maximum threshold, \mr{$N_j$}. The threshold is modulated by the word's corpus frequency, as frequent words generally require less processing than less frequent words, \mr{\label{r1-1.6}and word predictability. \mr{Note, however, that we did not include predictability effects in our model implementation. SWIFT also largely ignores low-level sensory perception and letter-level processing, which can have effects on the further (post-lexical) processing of a word and the sentence as a whole. In future work, processes such as bigram identification \citep{Snell2018} and surprisal \citep{Huang2023} are worth considering as extensions to SWIFT (or derivative models) to account for more aspects of lexical processing.}}

Once the word is identified, post-lexical processing begins and word activation decreases again. Post-lexical processing, however, is not explicitly modeled in SWIFT. Although SWIFT keeps track of the processing stage of words in the sentence, it has no higher-level representation of its constituents or of the entire word sequence. Adjacent words may have an influence on processing difficulty, but there is no mechanism to account for difficulty due to dependency completion processes at the sentence level.

While the relative word activations at the time of programming a saccade determine the relative probability of each word to be selected as the upcoming target, the timing of saccades is relatively independent \citep{Findlay1999} and involves a cascade of several processes. The cascade starts with a global timer, which triggers the \emph{labile} and subsequent \emph{non-labile} saccade stages, a distinction motivated by oculomotor performance in the double-step paradigm \citep{Becker1979}. During the labile stage, saccades can be canceled and a new target can be selected. During the non-labile stage, cancellation is no longer possible. The execution of the saccade itself is a noisy process subject to systematic (range) and random error \citep{McConkie1988}, where the systematic error component can be explained by a Bayesian-optimal estimation of the saccade target position \citep{Engbert2010}.

Target selection in SWIFT is inherently stochastic, as it depends on the dynamic, relative word activations at any given point in time. Words with high activation values are more likely to be selected as targets than words with lower activation. The probability $\pi_{j}\left(t\right)$ to select word $j$ at time $t$ as the next saccade target is given as 
\begin{equation}
\pi_{j}\left(t\right)=\frac{\left[a_{j}\left(t\right)\right]^{\gamma}}{\sum_{k=1}^{N_{\mathrm{W}}}\left[a_{k}\left(t\right)\right]^{\gamma}} \label{eq:selprob}
\end{equation}
where $N_{\mathrm{W}}$ is the number of words in the sentence and 
\mr{\begin{equation}
\label{eq:aj}
a_{j}\left(t\right)=\frac{n_j\left(t\right)}{N_\mathrm{a}}
\end{equation}}
is the \mr{normalized} activation of word $j$ at time $t$\mr{, which is the processing state of the word, normalized by parameter $N_\mathrm{a}$, the highest possible threshold of a word in a given corpus}.

The relation between the activation $a_{j}\left(t\right)$ of a word and its selection probability $\pi_{j}\left(t\right)$ also entails that words requiring little processing (i.e., ``easy-to-process'' words) \mrtwo{\label{r2-3.1}are less likely to be fixated, and thus more often skipped than less frequent (i.e., ``difficult-to-process'') words. This is because the former} pass through lexical and post-lexical processing faster \mr{and are} therefore in a state of higher activation for a shorter time period. The free parameter $\gamma$ modulates the relationship between word activations and selection probabilities. For $\gamma\rightarrow0$, words are selected randomly with equal probability, regardless of their actual activation values (if greater than zero). If $\gamma\rightarrow1$, there is a perfect linear relationship between activations and selection probabilities (Luce's choice rule). Higher values $\gamma\rightarrow\infty$ enforce a winner-takes-all principle so that the word with the highest activation always ``wins.''

\mr{\label{r1-2.7}The evolution of word activations in the original version of SWIFT \citep{Engbert2002,Engbert2005} was governed by ordinary differential equations (ODEs). In the more recent versions by \citet{Seelig2020,Rabe2021}, the dynamics of SWIFT changed toward a model with discrete internal states that evolve stochastically over continuous time.} Word activations and saccade timers are random walks that increase/decrease over time with different transition rates for different timers and individual word activations. The state of the model at time $t$ is given by a vector $n=(n_1,\,n_2,\,...,n_{4+N_\mathrm{W}})$, where the components $n_j$ represent the states of the subprocesses. \mr{\label{r1-2.3}Components} 1 to $N_\mathrm{W}$ are keeping track of the (post-)lexical processing of words, while \mr{components} $N_\mathrm{W}+1$ to $N_\mathrm{W}+4$ are saccade-related and additional stochastic variables (Table~\ref{Tab_States}). In each of the possible transitions from state $n=(n_1, n_2, ...)$ to $n'=(n'_1, n'_2, ...)$ only one of the sub-processes $n_i$ is changed by one unit.  The discrete stochastic variables $\{n_j\}$ at time $t$ map to the activation variables $\{a_j(t)\}$.

For the numerical simulation of the model, an algorithm can be derived from the \mr{master} equation \citep[see][for details]{Seelig2020},
\mr{
\begin{equation}
    \label{eq:master}
    \frac{\partial}{\partial t} p\left(n,t\mid n^{\prime\prime}\right)=\sum_{n'}\left[W_{nn'}p\left(n',t\mid n^{\prime\prime}\right)-W_{n' n}p\left(n,t\mid n^{\prime\prime}\right)\right]\;,
\end{equation}
which describes the temporal evolution of the model's internal states \citep{Gardiner1985,vanKampen1992}. It is specified by the transition rates $W_{n' n}$, which in turn govern the transitions between state vectors $n \mapsto n'$.}

\begin{table*}[!htbp]
\caption{Stochastic Transitions Between Internal States From $n=(n_1,n_2,...)\mapsto n'=(n'_1,n'_2,...)$}
\label{Tab_States}
\begin{center}
\begin{tabular}{lcrclcrcl}
\toprule
Process & & \multicolumn{3}{c}{Transition to \ldots} & & \multicolumn{3}{c}{Transition rate $W_{n' n}$} \\
\midrule
Word processing    & & $n'_j$&$=$&$n_{j}\pm 1$ & & \multicolumn{3}{l}{see Eq.~\eqref{eq:wword1} for $w_{j,\ldots,N_\mathrm{W}}$} \\ 
Saccade timer      & & $n'_{N_\mathrm{W}+1}$&$=$&$n_{N_\mathrm{W}+1}+1$ & & $w_{N_\mathrm{W}+1}$&$=$&$N_{t}/t_\mathrm{sac}\cdot (1+h\,a_k(t)/\alpha)^{-1}$  \\
Labile program     & & $n'_{N_\mathrm{W}+2}$&$=$&$n_{N_\mathrm{W}+2}+1$ & & $w_{N_\mathrm{W}+2}$&$=$&$N_\mathrm{l}/\tau_\mathrm{l}$ \\
Non-labile program & & $n'_{N_\mathrm{W}+3}$&$=$&$n_{N_\mathrm{W}+3}+1$ & & $w_{N_\mathrm{W}+3}$&$=$&$N_\mathrm{n}/\tau_\mathrm{n}$ \\
Saccade execution  & & $n'_{N_\mathrm{W}+4}$&$=$&$n_{N_\mathrm{W}+4}+1$ & & $w_{N_\mathrm{W}+4}$&$=$&$N_\mathrm{x}/\tau_\mathrm{x}$ \\
\bottomrule
\end{tabular}
\end{center}
\end{table*}%

Implementation of more detailed assumptions on the post-lexical stage can be achieved by changing the transitions rates $\{w_j(t)\}$ that control the stochastic transitions for the \mr{internal states $\{n_j(t)\}$ and thus for activations $\{a_j(t)\}$. \label{r1-3.8}Transition rates are a measure of the expected number of transitions in a given time unit (milliseconds in SWIFT) and are the inverse of the expected time between two consecutive transitions.}\footnote{\label{r2-3.2}\mrtwo{Given that the model parameters $\tau_\circ$ are defined as the expected total duration of the respective saccade stage, the transition rate for a single transition of a given saccade timer is the ratio of threshold and expected total duration.}} \mr{Transition rates, in combination with thresholds $N_j$, are therefore directly related to processing speed. While the rates for the saccade timers are either constant or determined according to an invariant rule (see Table~\ref{Tab_States}), the determination of transition rates for word processing components varies between processing stages, i.e.,}
\begin{align}
w_j\left(t\right) & =\begin{cases}
\alpha\cdot\Lambda_{j}\left(t\right) & \textrm{in lexical stage}\\
\max\left[\alpha\cdot\Lambda_{j}\left(t\right)\cdot\mathrm{proc},\omega\right] & \textrm{in post-lexical stage}\\
0 & \textrm{otherwise (complete)}
\end{cases}\;,\label{eq:wword1}
\end{align}
where $\alpha$ is the baseline processing difficulty, $\Lambda$ is the processing rate, $\textit{proc}$ is the relative processing speed for post-lexical processing, and $\omega$ is a minimum decay parameter.\footnote{The transition rate for post-lexical word $j$  cannot be lower than $\omega$, which ensures a decaying word activation even if there is no or little processing  at a given time $t$, e.g.\ when the word is not within the processing span.} In the integrated SEAM model, word activations in SWIFT are coupled with memory activations in LV05 in a Bayesian modeling framework by adapting the formula in Equation \eqref{eq:wword1}.

The fact that the SWIFT implements detailed mechanisms on word processing and saccade preparation is reflected by the number of parameters. Fitting the eye-movement model to experimental data started with hand-picking plausible parameter values, grid search \citep{Reichle1998}, genetic algorithms \citep{Engbert2002}, while optimizing the fit between empirical and simulated summary statistics. Based on the development of a likelihood approximation \citep{Seelig2020}, a fully Bayesian framework is now available for parameter inference \citep{Rabe2021}. The likelihood framework permits objective parameter fitting independent of a set of selected summary statistics, since fixation sequences are involved for likelihood computation. Using large-scale numerical simulations, it has been shown that SWIFT can reliably reproduce fixation durations, fixation probabilities and saccade amplitudes at the level of global and by-participant summary statistics, without using those summary statistics for the purpose of parameter fitting.

\subsection{SEAM: Activation-Based Coupling of SWIFT and LV05}
In baseline SWIFT, processing a word always starts out in the lexical processing stage. Once the word activation $n_{j}\left(t\right)$ has reached its threshold $N_{j}$ at time $t$, it begins post-lexical processing, and activation starts to decrease. When the activation has returned to zero, the word is completely processed.

\begin{figure}
\caption{Word Activation in SWIFT\label{fig:ex-word-act-swift}}

\includegraphics{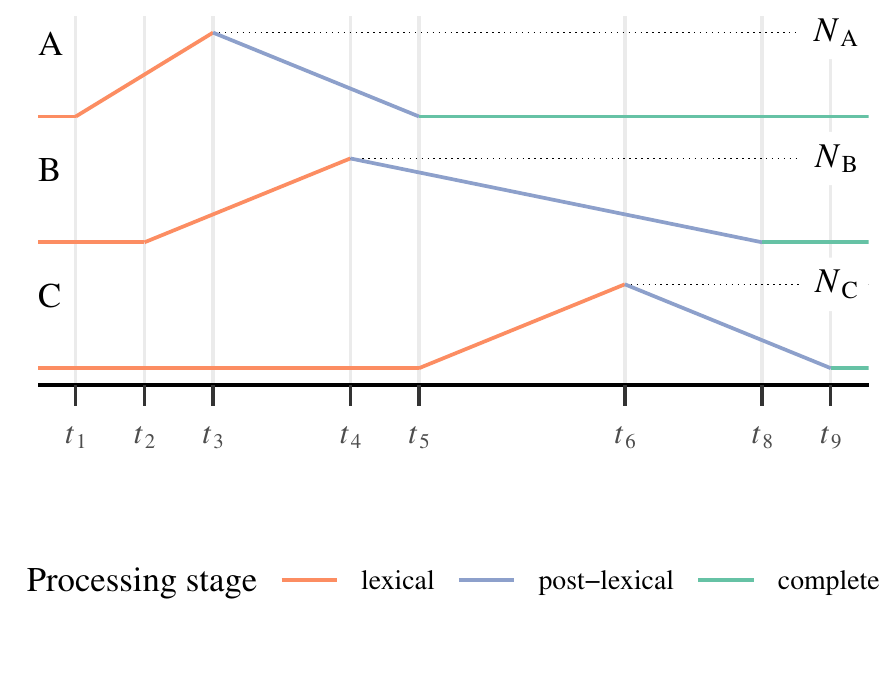}

\figurenote{Theoretical activation history of three words (A, B, and C). Colors of line segments correspond to the processing stage active at that given time. Activation maxima are $N_{\mathrm{A}}$, $N_{\mathrm{B}}$, and $N_{\mathrm{C}}$, respectively. Activations are displayed as continuous but are actually implemented as discrete counters.}
\end{figure}

Figure~\ref{fig:ex-word-act-swift} abstractly shows the activation histories of three hypothetical words. The figure assumes that the eyes move sequentially from word (a), to (b), to (c), leading to a somewhat sequential onset of their first processing ($t_{1}$, $t_{2}$, and $t_{5}$). The first stage of processing is the lexical stage.
During this stage, activations rise until they reach their respective maxima ($N_{\mathrm{A}}$, $N_{\mathrm{B}}$, and $N_{\mathrm{C}}$), which depend on printed word frequency. Given that saccade targeting depends on activation, the words in question are most likely to be selected as a saccade target if the upcoming saccade is programmed at times $t_{3}$, $t_{4}$, and $t_{6}$. This happens as well when the words enter the post-lexical processing stage. During post-lexical processing, activations decrease again, making it in turn less likely for the respective word to be selected as a target. Once the activation returns to zero ($t_{5}$, $t_{8}$, and $t_{9}$), the word is assumed to have completed processing.

A feature common to the SWIFT and LV05 is that both models use activation values to guide processing. SWIFT uses word activations to select words as saccade targets, while LV05 uses memory activations to select memory chunks as retrieval targets. Our integrated model SEAM keeps these activations separate, but implements an interaction, so that memory activations in LV05 modulate word activations in the SWIFT model. Therefore, rather than assuming that the sentence processor has direct control of the eye-movement targeting system, we propose an indirect, stochastic influence on saccade targeting via memory activations. This is in good agreement with eye-tracking studies carried out with larger-than-usual sample sizes that show that the effects of sentence processing cost \mr{\label{r1-0.3}due to memory interference} on \mrtwo{reading} measures have relatively small magnitudes  \citep[e.g.,][]{JaegerMertzenVanDykeVasishth2019}; \mr{larger} effect sizes are generally driven by \mr{lower-level} factors such as frequency and word length \citep{jemrsurprisal}. 

In SEAM, activations in the LV05 component reflect the construction of a sentence representation, which affect word activations and thereby stochastically influences target selection in the eye-movement component. As in SWIFT, the activation gradient of a word in SEAM is mainly determined by the transition rate, which varies between processing stages.
Compared to SWIFT, the sequence of processing stages in SEAM is extended by stages that reflect the cost of memory retrieval, which can account for \mr{additional} processing difficulty. \label{r1-1.7}Possible interactions of memory retrieval and the word activations \mr{during dependency resolution} include: (a)~\mr{the retrieval process delays} post-lexical processing of the \mr{the currently fixated region that caused the retrieval (that is, the retrieval trigger)}; and (b)~retrieval candidates are reactivated so that they attract regressions from \mr{the retrieval trigger}.

\begin{figure*}
\caption{Word Activation in SEAM\label{fig:ex-word-act-seam}}

\includegraphics{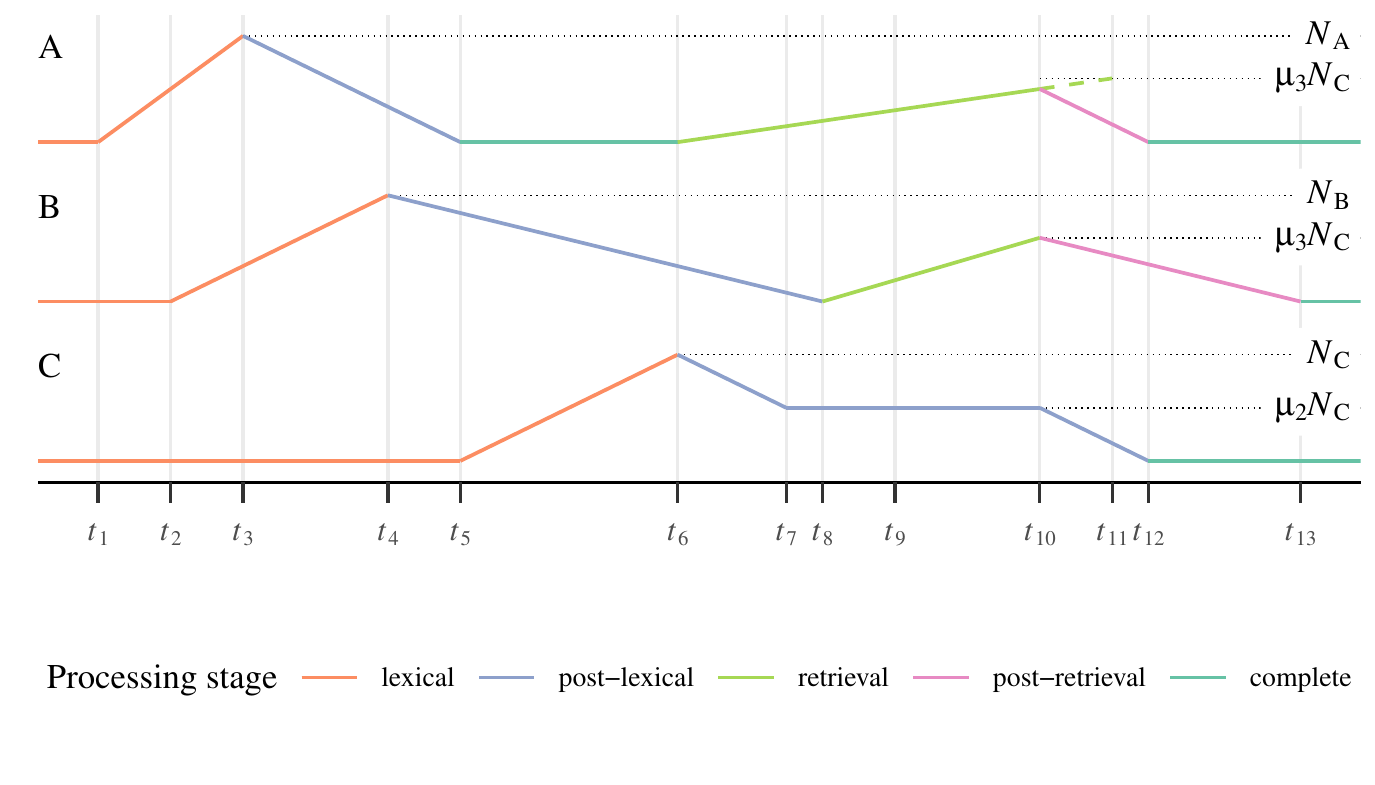}

\figurenote{Theoretical activation history of three words (A, B, and C). Colors of line segments correspond to the processing stage active at that given time. Activation maxima are $N_{\mathrm{A}}$, $N_{\mathrm{B}}$, and $N_{\mathrm{C}}$, respectively, for the transition from lexical to post-lexical processing, $\mu_{2}N_{\mathrm{C}}$ . Activations are displayed as continuous but are actually implemented as discrete counters.}
\end{figure*}

In Figure~\ref{fig:ex-word-act-seam}, activation histories of the same three words from the SWIFT example in Figure~\ref{fig:ex-word-act-swift} are shown. Like the baseline SWIFT model, words in SEAM go through a lexical and post-lexical processing stage before they are considered \emph{completely processed.} However, SEAM additionally accounts for the resolution of a linguistic dependency during post-lexical processing of word~C. Once the words are lexically accessed ($t_{3}$, $t_{4}$, and $t_{6}$), they are encoded as chunks in SEAM's memory module, along with their features, as in the LV05 model. Words~A and B are assumed to not trigger a dependency completion process; this is the case for most nouns. However, when word~C, which could be a verb, is processed and the associated chunk is stored in memory, a subject-verb dependency must be resolved. A retrieval is thus triggered. The assumption that nouns do not trigger a dependency completion process is obviously an oversimplification; but this simplification is reasonable for the data being modeled in this paper, as in the experiment design of \citet{Mertz21}, the theoretically interesting dependency completion occurs at the verb.

During retrieval, all words that are fully processed before the processing of word~C completes are counted as retrieval candidates. Candidate words enter into a retrieval stage in which activation increases until the retrieval process finishes.\footnote{\label{r1-3.5}A word can also become a candidate after the retrieval process has started. Word A, for example, is already a candidate at the time the post-lexical processing of word C starts at time $t_{6}$, given that it was already completely processed at time $t_{5}$. Therefore, the retrieval stage of word A starts immediately with the start of the post-lexical stage of word C. \mr{This mechanism is necessary for the rare but possible case that a retrieval is triggered before any candidate word has been encoded as a chunk in memory.}} The activation increase differs by the degree to which the retrieval candidate features match the retrieval cues, implementing a core assumption of the LV05 model.

\mr{\label{r1-1.8}The effect of the memory activations on word activations is mainly modulated by the new parameters $\mu_2$ and $\mu_3$. }The retrieval stage ends when one candidate reaches a threshold value, which is a fraction $\mu_{3}$ of the maximum activation of the retrieval trigger $N_{\mathrm{C}}$. Because post-lexical processing in SEAM is only finished after all dependencies have been resolved, the post-lexical activation of the retrieval trigger is guaranteed not to fall below a fraction $\mu_{2}$ of its maximum activation during retrieval. This is why the post-lexical activation of word~C does not change between $t_{7}$ and $t_{10}$. In this example, despite entering the retrieval phase at a later time, word~B reaches the retrieval threshold at time $t_{10}$ before word~A, thereby concluding the retrieval process. Consequently, the post-lexical processing of word~C continues and all retrieval candidates, that is, word~A and word~B, enter a post-retrieval stage, which is equivalent to an additional post-lexical processing stage. This also entails that the retrieval phase of word~A is aborted, which would otherwise have reached threshold at time $t_{11}$.

The transition rates of the baseline SWIFT model for word $j$, Equation~\eqref{eq:wword1},  are replaced by
\begin{equation}
w_j'\left(t\right)=\begin{cases}
\alpha\cdot\Lambda_{j}\left(t\right) & \textrm{in lexical stage}\\
\max\left[\alpha\cdot\Lambda_{j}\left(t\right)\cdot\mathrm{proc},\omega\right] & \textrm{as retrieval trigger (}j=m\land n_{j}\left(t\right)>\mu_{2}N_{j}\textrm{)}\\
0 & \textrm{as retrieval trigger (}j=m\land n_{j}\left(t\right)\leq\mu_{2}N_{j}\textrm{)}\\
\max\left[\alpha\cdot\Lambda_{j}\left(t\right)\cdot\mathrm{proc},\omega\right] & \textrm{in post-lexical stage}\\
\mu_{3}N_{m}F^{-1}\exp\left[A_{j,m}'\left(t\right)\right] & \textrm{as retrieval candidate (}j\neq m\textrm{)}\\
\max\left[\alpha\cdot\Lambda_{j}\left(t\right)\cdot\mathrm{proc},\omega\right] & \textrm{in post-retrieval stage}\\
0 & \textrm{otherwise (complete)}
\end{cases}\;\label{eq:wword-seam}
\end{equation}
where $m$ is the current retrieval trigger that needs to form a dependency. \mr{\label{r1-3.3}The transition rate for the retrieval candidate $j$, triggered by dependency resolution for word $m$, is chosen to ensure that the total duration of reaching threshold (i.e., the time for $j$ to be matched as a dependent of $m$), matches the retrieval latency predicted by LV05. Therefore, it is computed as the threshold value $\mu_3N_m$ divided by the expected total duration of $j$ in that stage, $F\exp\left[-A_{j,m}'\left(t\right)\right]$.}

Altogether, SEAM extends the baseline SWIFT model parameters \citep{Rabe2021,Seelig2020} with seven additional model parameters. The parameters $d$ (decay), $S_{\mathrm{max}}$ (maximum memory activation strength), $F$  (retrieval latency scaling factor) and $p$ (mismatch penalty), which modulate $w'\left(t\right)$ through $A_{j,m}'\left(t\right)$, are directly based off their LV05 implementations \citep{Lewis2005}. Moreover, the link between word activations in LV05 and processing rate in SWIFT is complemented by the three new model parameters $\mu_{1}$, $\mu_{2}$, and $\mu_{3}$, as detailed \mr{above}. Some parameters of the LV05 model, in particular for goal activation and noise ($G$, and $\mathit{ans}$), are ignored in the present implementation. Variation in the goal activation parameter is usually used to model individual-level capacity differences \citep[e.g.,][]{DailyEtAl2001,MaetzigEtAlTopiCS,VasishthEngelmann2020}; \mrtwo{these are not being investigated in the present work}. The goal activation is fixed at 1.0, which gives equal weight to all retrieval cues. The noise parameter $\mathit{ans}$ is replaced by the built-in stochasticity of SWIFT. Moreover, the parameters $S_\mathrm{max}$ and $F$  are not independent in terms of the resulting \mr{\label{r1-3.9}retrieval latency and transition rate}, which is why we will only estimate $F$ as a free parameter and keep $S_\mathrm{max}$ at a fixed default value of $1.5$. In the present study, we also exclude $\mu_1$, the fixed time needed to execute a production rule, by setting it to $0$, because we assume this time to overlap with some of the oculomotor processes already present in the model. Since $S_\mathrm{max}$ is fixed, we also decided to fix mismatch penalty $p$ at its default value, as the relation of the two parameters is critical. Thus, the only parameters that were fit to the \citet{Mertz21} data were $F$, $d$, $\mu_2$, and $\mu_3$. For a complete list of model parameters and default values in SEAM, see Appendix~\ref{app:parameters}.

For our implementation of SEAM, we opted for a simplified version of the LV05 model \citep{Engelmann2015} and the latest version of SWIFT \citep{Rabe2021}.\footnote{The principal reason for using the simplified version of the LV05 model is tractability. Using the full ACT-R architecture, which is {Lisp}-based, would require much more complex engineering decisions, and would make the model inaccessible to researchers unfamiliar with Lisp but who are interested in exploring its behavior with novel data.} SEAM connects the baseline eye-movement control architecture of SWIFT with the interactive working memory module of LV05 via activation-based coupling: reading words in SWIFT leads to the creation of memory chunks and can trigger retrievals in LV05, whereas chunk activations computed by LV05 modulate word activations in SWIFT. 

\section{Data Availability}
All experimental and simulated data, analysis code, and computational models (SEAM and SWIFT) reported in this paper are available at the Open Science Framework \mrtwo{(\url{https://doi.org/10.17605/OSF.IO/8ZRXB})} and at \mrtwo{the University of Potsdam (\url{https://gitup.uni-potsdam.de/CRC1294/B03/SEAM-2023-Paper})}.

\section{Experimental Study (Mertzen et al., 2023)}
To test the predictions of the integrated model, we use data from a memory interference experiment conducted with 61 English native speakers \citep{Mertz21}. This experiment was originally planned with 120 participants, but due to the pandemic, data collection had to be aborted. Our inability to reach the target number of participants has consequences for model evaluation, as discussed later.

The \citet{Mertz21} experiment employed a fully crossed \mr{distractor} subjecthood~(2) $\times$ animacy~(2) design that closely mirrored an experiment reported in \citet{VanDyke2007}. Examples of the four conditions are shown below in example (\ref{ex2}).

\pex\label{ex2}
\a \label{ex2a}It turned out that the\textbf{ attorney}$_{+\mathit{anim}}^{+\mathit{subj}}$ whose secretary had forgotten about the important \uline{meeting}$_{-\mathit{anim}}^{-\mathit{subj}}$ frequently \textbf{complained}$\left\{ _{\mathit{anim}}^{\mathit{subj}}\right\} $ about the salary at the firm.
\a  \label{ex2b}It turned out that the\textbf{ attorney}$_{+\mathit{anim}}^{+\mathit{subj}}$ whose secretary had forgotten about the important \uline{visitor}$_{+\mathit{anim}}^{-\mathit{subj}}$ frequently \textbf{complained}$\left\{ _{\mathit{anim}}^{\mathit{subj}}\right\} $ about the salary at the firm.
\a  \label{ex2c}It turned out that the\textbf{ attorney}$_{+\mathit{anim}}^{+\mathit{subj}}$ whose secretary had forgotten that the \uline{meeting}$_{-\mathit{anim}}^{+\mathit{subj}}$ was important frequently \textbf{complained}$\left\{ _{\mathit{anim}}^{\mathit{subj}}\right\} $ about the salary at the firm.
\a \label{ex2d}It turned out that the\textbf{ attorney}$_{+\mathit{anim}}^{+\mathit{subj}}$ whose secretary had forgotten that the \uline{visitor}$_{+\mathit{anim}}^{+\mathit{subj}}$ was important frequently \textbf{complained}$\left\{ _{\mathit{anim}}^{\mathit{subj}}\right\} $ about the salary at the firm.
\xe

In the example above, processing the verb \emph{complained} is expected to trigger a retrieval for an animate subject noun phrase. In all sentences, \emph{attorney} is the grammatically correct subject of \emph{complained}, and should thus be retrieved. However, the distractor noun phrase (\emph{meeting} or \emph{visitor}) may interfere with the retrieval of \emph{attorney}. The distractor is \emph{visitor} in the $+\mathrm{animate}$ or \emph{meeting} in the $-\mathrm{animate}$ condition, and it is either a subject ($+\mathrm{subject}$) or an object ($-\mathrm{subject}$) of the embedded clause.

According to cue-based retrieval theory, both subjecthood and animacy of the distractor should lead \mr{to} additional difficulty for resolving the critical dependency. This is due to the fan effect \citep[e.g.,][]{Anderson1990}, which is also known as similarity-based interference \citep{JaegerEngelmannVasishth2017}:  When the feature specification of a distractor overlaps with that of the retrieval target, it diverts some of the retrieval activation from the target to itself. The activation of both the target and distractor are reduced, leading to longer retrieval time; what ends up being retrieved in a particular simulation run (target or distractor) depends on which chunk happens to have higher activation (this can vary in simulation runs due to stochastic noise in the activation). It is therefore possible that the distractor is sometimes erroneously retrieved. As indices of increased processing difficulty, we expect additive effects of animacy and subjecthood of the distractor on regression path duration and outgoing regression probabilities on the critical verb (\emph{complained}). The primary region of interest where the effect of the subjecthood and animacy manipulation should manifest is the verb; however, because similarity-based interference effects have been shown to occur in the region just before the verb \citep{VanDyke2007, Lago2021}, \citet{Mertz21} also investigated the effect at the adverb (\emph{frequently}) that preceded the critical verb. For this reason, in our investigations we also report model fits for this pre-critical region.
 
In summary, similarity-based interference accounts predict that conditions (\ref{ex2}b,d) should be more difficult to process than conditions (\ref{ex2}a,c) due to the animacy of \emph{visitor}, and conditions (\ref{ex2}c,d) should be more difficult to process than conditions (\ref{ex2}a,b) due to the distractor being in subject position.

As indices of increased processing difficulty, additive effects of distractor animacy and distractor subjecthood
were expected in reading times and outgoing regression probabilities. An interaction of distractor subjecthood and animacy was not predicted but is reported in \citet{Mertz21}  for completeness; in the \citet{Mertz21} analysis, there was no evidence for an interaction.

In this summary of the \citet{Mertz21} results, we report only regression path duration and outgoing regression probabilities \mr{\label{r1-1.11}(FPR\mrtwo{, also referred to as \emph{first-pass regressions out} in the original study})} from the pre-critical adverb and the critical verb; for full details of all experimental results, please see the original paper.

The effects of animacy and subjecthood (coded as sum contrasts) were analyzed using Bayesian mixed-effects models. Subject and item were specified as random effects in the models, with a full variance-covariance matrix for subject and item random effects.  The models were implemented with \textit{brms} \citep{Buerkner2017,Buerkner2018,Buerkner2021}, an interface to \textit{Stan} \citep{Carpenter2017}. Priors were mildly informative Gaussian distributions for the linear model coefficients (intercept and slopes) and mildly informative regularizing Lewandowski-Kurowicka-Joe (LKJ) priors \citep{lewandowski2009generating} for random effects correlation matrices; setting the LKJ prior's parameter $\nu$ to $2$ downweights extreme correlations like $\pm 1$. For a detailed tutorial on linear mixed models in the Bayesian setting, see chapter 5
of \citet{NicenboimEtAlBayes2019}, or \citet{SorensenVasishthTutorial}. 

\begin{figure}
\caption{Experimental Effects of \citet{Mertz21}\label{fig:experimental-effects}}

\includegraphics{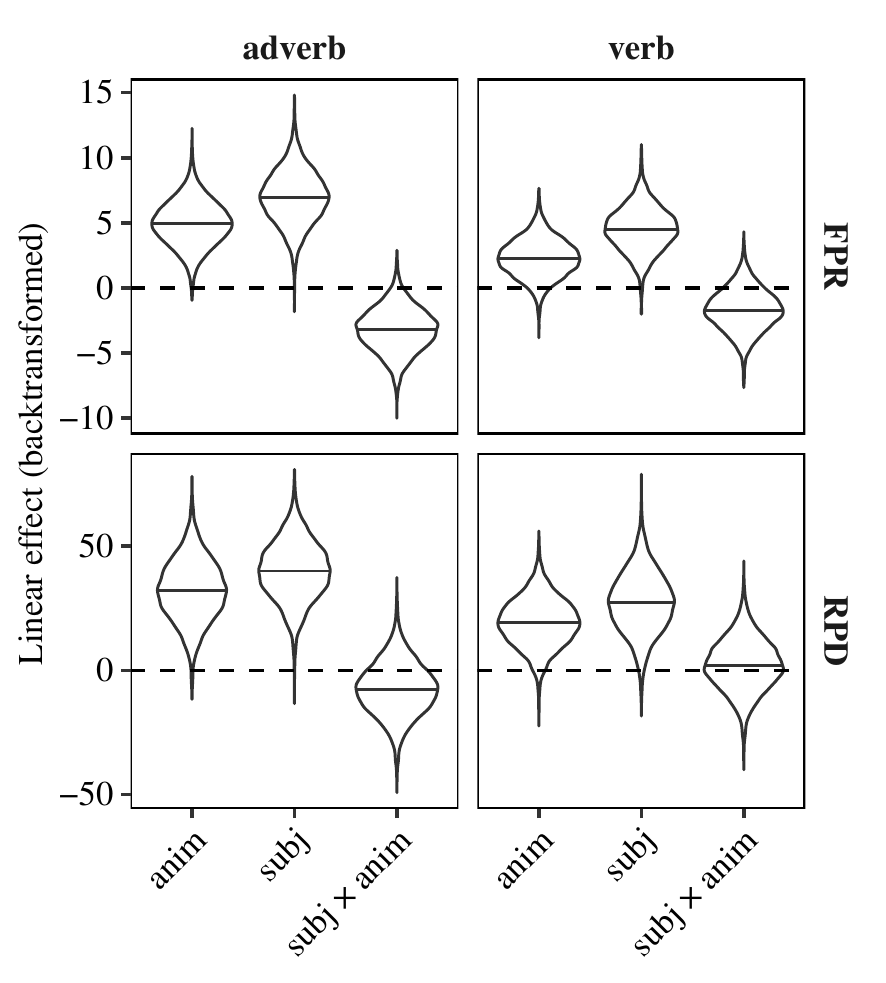}

\figurenote{Plotted violins are the estimated posterior distributions of experimental effects of subjecthood (\emph{subj}) and animacy (\emph{anim}) on regression-path duration (RPD) and first-pass outgoing regression probability (FPR) from Bayesian mixed-effects regressions, as analyzed and reported by \citet{Mertz21}. Posteriors are backtransformed linear effects in ms (for RPD) or \% (for FPR).}
\end{figure}

The results in \citet{Mertz21} showed reading time patterns consistent with effects of subjecthood (syntactic interference) and effects of animacy (semantic interference). 
Figure~\ref{fig:experimental-effects} shows that on the pre-critical adverb, the effect of subjecthood shows longer regression path duration (RPD) and \mrtwo{higher outgoing regression probability (FPR)} for conditions that have a +subject distractor (95\% credible intervals (CrIs): RPD $[17, 63]$~ms, FPR $[3, 11]$\%). Similarly, the effect of animacy shows longer regression-path duration and an increase in \mrtwo{outgoing regression probability} for conditions with animate distractors compared to conditions with inanimate distractors (95\% CrIs: RPD $[8, 57]$~ms, FPR $[2, 8]$\%). The subjecthood $\times$ animacy interaction in regression-path duration is centered on zero; for first-pass regressions, the interaction has a negative sign ($[-7, 0]$\%).

On the critical verb, the effects of subjecthood and animacy show a similar pattern of longer regression path duration and an increase in \mrtwo{outgoing regression probability} (Subjecthood 95\% CrIs: RPD $[3, 52]$~ms, FPR $[1, 8]$\%; Animacy 95\% CrIs: RPD $[0, 39]$ ms, FPR $[-1, 5]$\%). The interaction is centered around zero for regression path duration and \mrtwo{outgoing regression probability}.
The increased reading times and regressions for conditions that have subject or animate distractors indicate that syntactically and semantically similar distractors can interfere during long-distance dependency formation.  

\section{Simulation Study}
\label{r1-2.10-bw1}The reliability of computational cognitive models critically depends on the availability of appropriate methods for statistical inference \citep{Engbert2022,Schuett2017}.  We previously applied a broader principled Bayesian workflow \citep{Schad2020} for the baseline SWIFT model in \citet{Rabe2021}, which is used as the eye-movement platform in SEAM.

\begin{figure*}
\caption{Example Simulation in SWIFT\label{fig:ex-sim-swift}}

\includegraphics{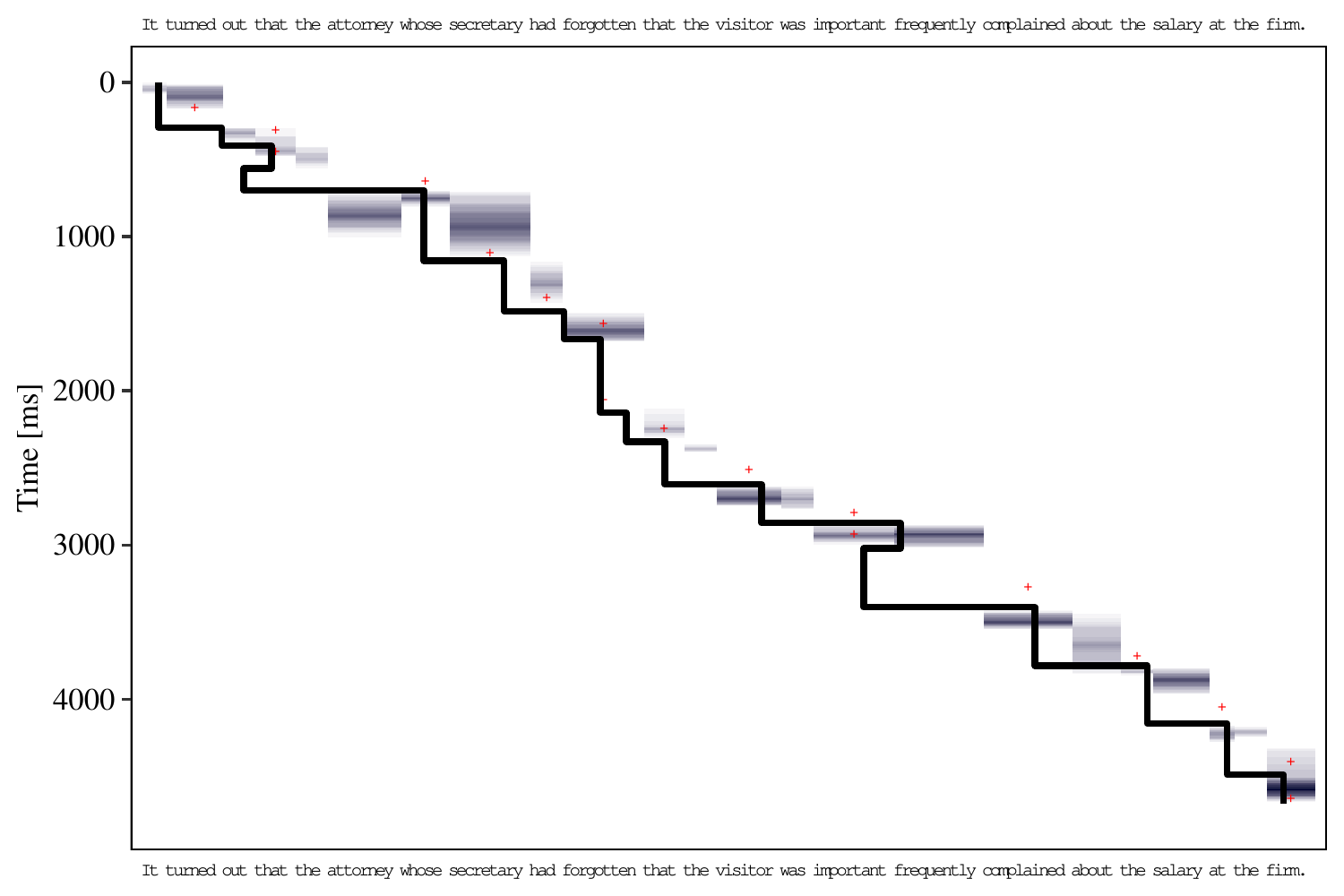}

\figurenote{\mr{SWIFT simulation for Example (\ref{ex1}). The bold black line is the simulated fixation location (x-axis) as a function of time (y-axis). Saccades are horizontal displacements of the black line. Word activations are depicted by gradients in the background, with darker shades referring to higher activation. The target selection preceding each executed saccade is depicted by a red cross, marking both the time and intended saccade target. Target selection is based on the relative word activations at the respective time point of saccade programming. Saccade timers, which are also components of the internal states, are omitted for brevity. For more details, see \citet{Seelig2020,Rabe2021}.}}
\end{figure*}

\begin{figure*}
\caption{Example Simulation in SEAM\label{fig:ex-sim-seam}}

\includegraphics{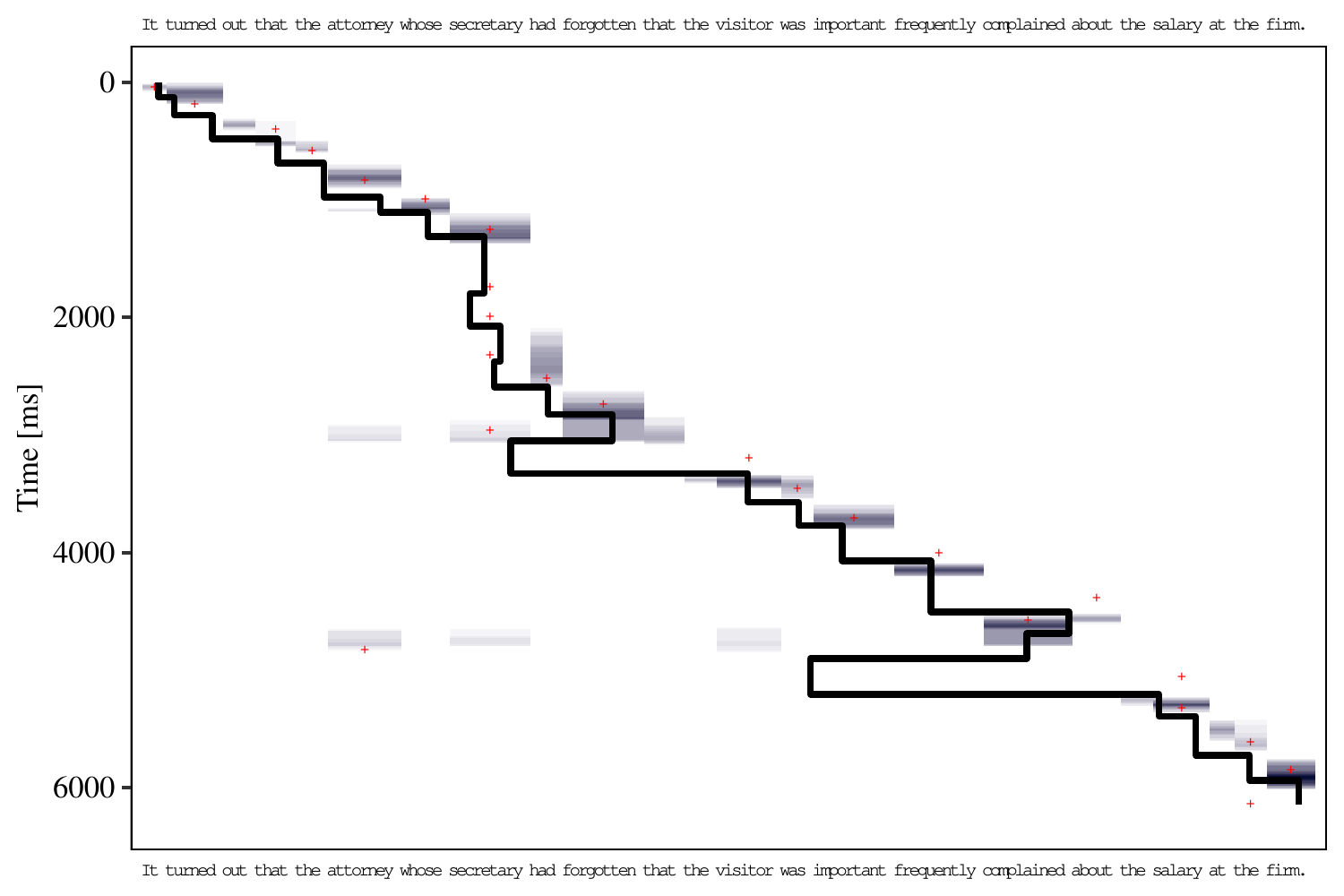}

\figurenote{\mr{SEAM simulation for Example (\ref{ex1}). As in Figure~\ref{fig:ex-sim-swift}, the black line is the simulated fixation location (x-axis) as a function of time (y-axis), gradients in the background are word activations, and red crosses are selected targets. Note that, in comparison to Figure~\ref{fig:ex-sim-swift}, processing of \emph{forgotten} and \emph{complained} triggers retrievals, which prolongs processing of the trigger and reactivates potential retrieval candidates. In this simulation, during both retrievals, regressive saccades are programmed and executed.}}
\end{figure*}

\mr{\label{r1-3.1}In Figures~\ref{fig:ex-sim-swift} and \ref{fig:ex-sim-seam}, we visualized the word activation field and eye trajectory for a simulated trial in SWIFT and SEAM, respectively. As can be seen, SEAM behaves similarly to SWIFT throughout most of the trial. However, the models' behaviors start to diverge when the verb \emph{complained} is processed and triggers a retrieval in SEAM. During the retrieval phase, word activations of previous words that have been encoded as memory chunks increase. Words with better cue match for the retrieval approach the activation threshold faster than those with lower cue match. If a saccade is triggered during the retrieval phase, the reactivated words can attract regressions.}

Without proper checks, it is not self-evident that Bayesian model fitting of SEAM can be carried out in the same way as for SWIFT. However, we expect that our implementation of SEAM will exhibit correct inference because it meets the following three critical conditions: First, for all observables that were taken into account (i.e., fixation positions and durations), a model likelihood has already been implemented in SWIFT \citep{Seelig2020}. Secondly, both SWIFT and LV05 are dynamic in the sense that they describe activation values as a function of time, which allows us to let them interact dynamically without a significant modification of their initial conceptualization. Thirdly, the dynamics of eye movements and sentence processing interact in the integrated SEAM model and will thus affect the observable temporal and spatial aspects of fixation sequences due to the activation coupling of the constituent SWIFT and LV05 components. The coupling via word activations permits indirect fitting of model parameters related to memory retrieval, as long as they have some probabilistic effect on the outcome variables captured by SWIFT.

Given these properties, we tested the computational faithfulness of SEAM using the Markov Chain Monte Carlo (MCMC) sampling algorithm DREAM$_{\text{zs}}$ \citep{Laloy2012} based on profile log-likelihoods and model parameter recovery, similar to the approach taken in \citet{Rabe2021}. The DREAM$_{\text{zs}}$ \citep{Laloy2012,terBraak2008,Vrugt2009} sampler has previously been successfully used with complex dynamical models of eye-movement control, including SWIFT for reading \citep{Rabe2021} and SceneWalk for scene viewing \citep{Schwetlick2020,Schwetlick2022}. 

After confirming the computational faithfulness of the model, we fitted the model to a training subset of the experimental data and compared predictions for a withheld test portion using relevant global summary statistics and the predicted experimental effects of similarity-based interference\footnote{\mr{\label{r1-3.11}We will focus on effects of subjecthood and animacy cues, since those were of main interest in the experimental study. Additional features/cues such as clause locality were also encoded but are not of primary interest here. The full memory and retrieval schedules are available in the model supplement at \url{https://github.com/mmrabe/SEAM-2023-Paper/tree/main/SEAM/DATA}.}} described in the previous section.

\subsection{Method}
\subsubsection{Data Assimilation}
\label{r1-2.10-bw2}In eye-movement research, the experimental (observed) data are fixation sequences consisting of time-ordered sequential observations. In such a case, the identification of model parameters is possible within the field of \emph{data assimilation} \citep{Reich2015,Engbert2022}. Data assimilation refers to the integration of complex mathematical models with time-series data \citep[see][for an introduction]{Morz18}. In this framework, the SWIFT model has previously been implemented for Bayesian model fitting \citep{Seelig2020}. \citet{Rabe2021} showed that, in a principled Bayesian workflow \citep{Schad2020}, SWIFT can be reliably fitted to simulated and experimental data even with many free parameters and sparse data that resulted from splitting by participant and experimental condition. 

\subsubsection{Sequential Likelihood}
The time-ordered nature of fixational eye movements make them a suitable target for data assimilation \citep{Engbert2022}. To exploit the sequential information of the data, some of those models use \emph{sequential likelihoods} for parameters $\boldsymbol{\theta} \in \boldsymbol{\Theta}$ such that

\begin{equation}
    L_{M}\left(\boldsymbol{\theta} \mid X_n\right) = \mr{P_{M}\left(x_1 \mid \boldsymbol{\theta}\right) \prod_{i=2}^n P_{M}\left(x_i \mid X_{i-1}, \boldsymbol{\theta}\right)}\;,
\end{equation}
where $X_n=(x_1,\ldots,x_n)$ is the entire sequence of $n$ events and \mr{$P_{M}\left(x_i \mid X_{i-1}, \boldsymbol{\theta}\right)$} is the likelihood of the $i$-th event of the sequence given all previous events $X_{i-1}=(x_1,\ldots,x_{i-1})$.

Successful examples of applying data assimilation for visual tasks are, for example, SceneWalk \citep{Schwetlick2020,Schwetlick2022} for scene viewing and SWIFT \citep{Seelig2020,Rabe2021} for reading. There, each event of the sequence, $x_i$, is a fixation. Since the location \mr{and temporal onset} of the first fixation \mr{are} typically known due to the experimental paradigm, e.g., sequences always starting at a fixation cross, the likelihood for $x_1$ is given by $ \mr{P_{M}\left(x_1 \mid\boldsymbol{\theta}\right)}=1$. SceneWalk and SWIFT further decompose the likelihood into spatial and temporal components, since each fixation has a spatial location on the screen and a duration.

As SEAM is based on SWIFT and we only changed the latent transition rates rather than the saccade execution itself, we can easily use the data assimilation methods implemented for SWIFT. This is especially useful because we fit the model on a by-participant basis and hence only have little data for parameter estimation. The decomposition of temporal and spatial likelihood components is also theoretically interesting since we can expect the modification of the transition rates to affect both the temporal control and target selection of the (simulated) saccadic eye movements.

\subsubsection{Profile Likelihoods}
As SEAM modifies model dynamics and thus the likelihood function of SWIFT, a reevaluation of the \emph{profile log-likelihoods} is crucial. Those are generated by first simulating data with known parameter values, and then systematically varying parameter values and inspecting the likelihood of the data for each value. Ideally, the likelihood of the data should be highest for the true parameter values. In order to assess whether the modifications introduced in SEAM are appropriately captured in its likelihood, it should be ensured that the newly introduced free parameters affect the outcome likelihood. Thus, the behavior of the likelihood as a function of each of the new parameters represents a necessary condition for identifiability and statistical inference of the full model \citep{Seelig2020,Rabe2021}. 

Parameters were inspected if they were going to be fitted later on and/or were added in this model implementation compared to the reference SWIFT implementation \citep{Rabe2021}. This was the case for a total of 11 parameters (see Figure~\ref{fig:likelihood-profiles}). Parameters $\mu_{1}$ and $S_{\mathrm{max}}$ were also inspected even though they were not selected to be fitted to the recovery and experimental data. This is because the parameters themselves are identifiable, as can be seen in Figure~\ref{fig:likelihood-profiles}, but they are not independent from other model parameters in terms of an effect on model behavior. All other shown model parameters are also fitted to simulated data for parameter recovery as well as to experimental data.

\subsubsection{Parameter Estimation and Recovery}
As a last step for the verification of the computational faithfulness of the approach, we applied a sampling algorithm to simulated data with known true parameter values in order to ensure the validity of the computational approach. We generated 100 unique data sets with different sets of true parameters $\boldsymbol{\theta}^\star$ randomly sampled from the prior distribution later used for parameter estimation. Parameters would be considered successfully recovered if the correlation between true and recovered parameters was sufficiently high and the normalized root mean squared error (NRMSE) was sufficiently low.

\subsubsection{Summary Statistics and Experimental Effects} 
Even though we are using an objective likelihood-based approach for model fitting, it is important that simulated and empirical data are in good agreement at the level of relevant summary statistics, especially with regard to comparability with competitor models and theory testing \citep{Roberts2000}. Because the goal for SEAM is to explain both spatial and temporal aspects of eye movements in reading, we consider a number of different spatial and temporal summary statistics frequently used in reading research. For the spatial dimension, we are looking at several fixation probabilities, that is, probabilities to fixate (or skip) specific words under different conditions. For the quantification of the temporal aspects of the model fit, we evaluate different fixation durations, that is, average reading times under different conditions.

A subset of the experimental test data set is withheld from parameter estimation, and this held-out set will then \mr{be} compared on the basis of summary statistics against predicted data from SEAM and SWIFT using estimated parameters. Specifically, we first split the experimental data into a training and test subset, fitting the model to 70\% of the data (training set) of each participant and condition, subsequently predicting eye trajectories for the other 30\% (test set). For each withheld trial, we generated a fixation sequence using the HPDI (highest posterior density interval) midpoint of the sampled posterior distribution of a given participant and parameter \citep{Rabe2021}. We also present the predictions of SEAM and SWIFT for the experimental memory interference effects, which can be similarly derived from the simulated and experimental data alike.

\subsection{Results}
\subsubsection{Profile Likelihoods}
We evaluated the likelihood for a typically sized simulated data set where all parameters had been set to default values\footnote{\mr{\label{r1-3.2}These values vary slightly from the defaults used in  \citet{Seelig2020,Rabe2021}. They are not to be understood as universally valid defaults but as fixed values wherever they are not fitted, and are merely reported here for reasons of transparency and reproducibility.}}  (see Appendix~\ref{app:parameters}). For each parameter, the respective true value, that is, the value used for simulating the data set, is shown with a vertical dashed red line. Then, for each parameter, for 50 equidistant parameter values in the intervals shown, the likelihood for the data given the model was evaluated. Ideally, the likelihood should be maximal around the true value. 

\begin{figure*}
\caption{Example Profile Log-likelihoods\label{fig:likelihood-profiles}}

\includegraphics{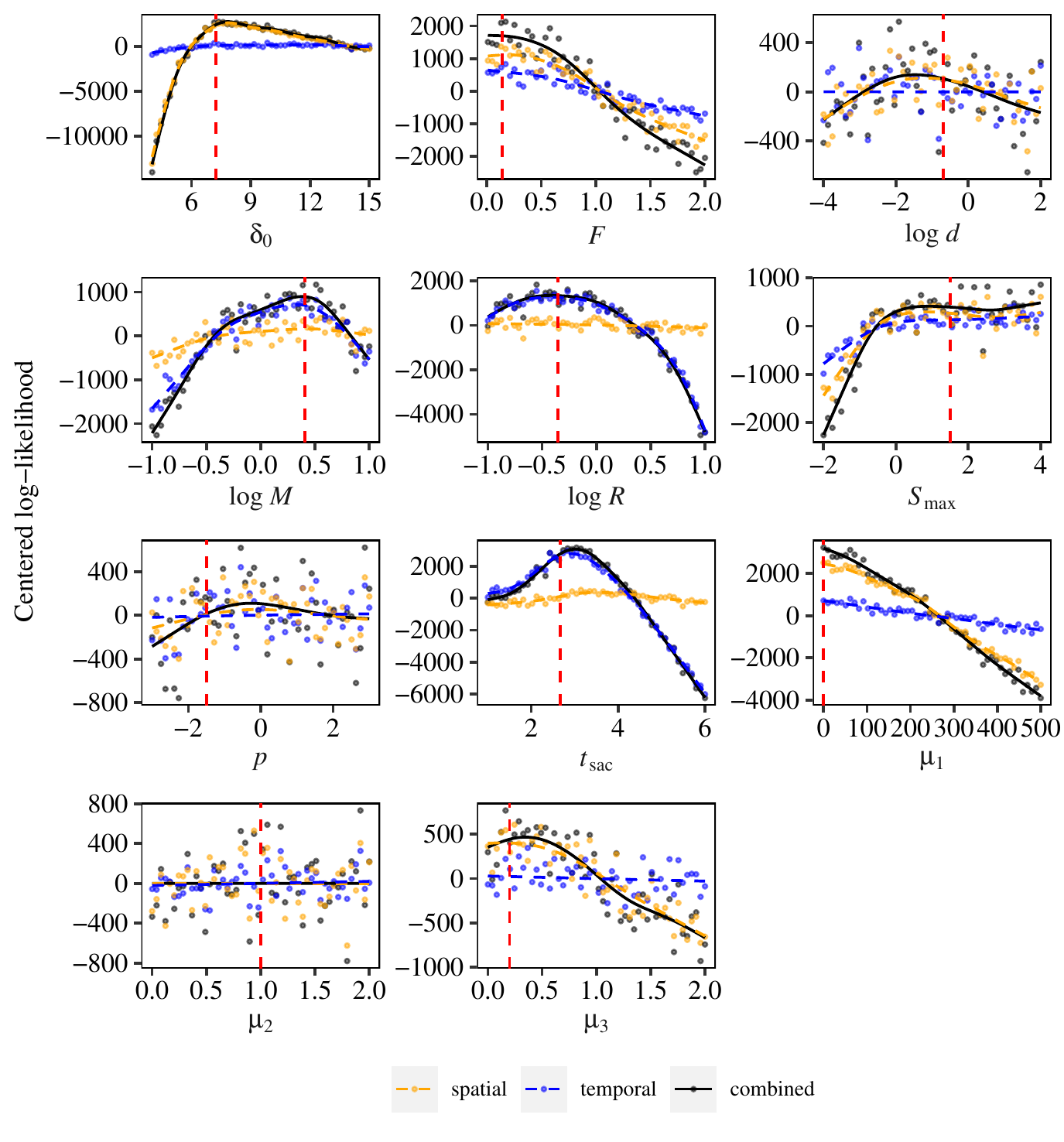}

\figurenote{Centered profile log-likelihoods $\log L_{M}\left(\boldsymbol{\theta}\mid X\right)$ for a simulated data set $X$ with known/true parameters $\theta^{\star}$. Profiles are generated by varying one parameter (dimension) of $\boldsymbol{\theta}$ at a time while holding the others constant at their respective \emph{true} parameter value. True parameter values are denoted by the vertical red line. Dots in the background are individual stochastic pseudo-likelihood
evaluations, each with a spatial and temporal likelihood component, and their combination (sum). Curves are GAM smooths on those individual evaluations.}
\end{figure*}

In Figure~\ref{fig:likelihood-profiles} we observe that the likelihood peaks, as expected, around the true value for most of the parameters. This means that (i) the parameters affect the likelihood and (ii) the likelihood may be used to recover their values. Individual likelihood evaluations are represented by dots. The plotted line smooths are just for guidance and do not represent the true likelihoods. The important observation here is that the highest evaluated likelihoods are always relatively close to the true value, even for the case of $\mu_{2}$, where the smoothed lines falsely suggest a flat likelihood.

Since not every fixation involves a retrieval, the new SEAM parameters can only have a very limited effect on the likelihood. Therefore, effects observed in the likelihood function are less pronounced than for the established SWIFT parameters such as processing span $\delta_{0}$.  The fact that that higher likelihood evaluations nevertheless cluster around the true values is an indication that the parameters are identifiable, but their fitted values should be interpreted with caution.

\mr{\label{r1-3.13} For one of the parameters, $\mu_2$, the likelihood does not peak at all, which is probably because $\mu_2$ only affects the model's behavior in rare instances. As $\mu_2$ only determines the threshold value of a retrieval trigger, the likelihood is only affected for the small subset of words that trigger retrievals. By contrast, $\mu_3$ affects the threshold of multiple words at the same time, i.e., all words previously processed. Also note that the profile likelihoods as well as the parameter recovery reported below are based on simulated data sets comparable in size to the experimental data of \citet{Mertz21}. We would expect $\mu_2$ to exhibit a more pronounced effect on the likelihood for larger data sets with more retrieval events. Despite the noise level of the profile likelihood of $\mu_2$, we decided to fit $\mu_2$ as a free parameter. This means that different plausible values from the prior are considered throughout the sampling procedure instead of keeping $\mu_2$ fixed at a (possibly implausible) default value.}

\subsubsection{Parameter Recovery}
\begin{figure*}
\caption{Parameter Recovery of SEAM Parameters\label{fig:recovery}}

\includegraphics{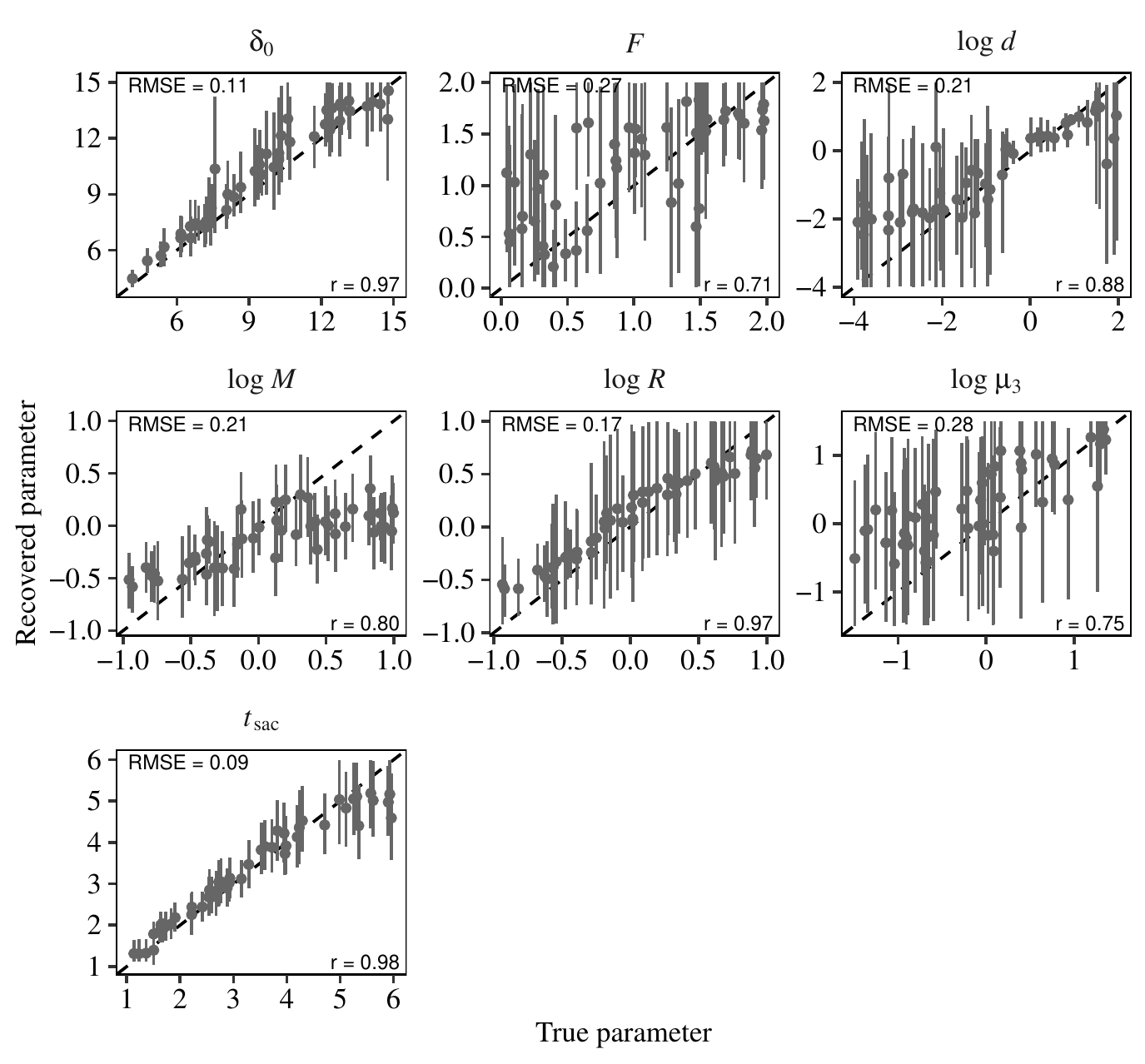}

\figurenote{Results of a parameter recovery for 50 simulated data sets, for which the parameters were randomly drawn from a uniform distribution with the bounds shown on the x-axes. 95\% credible intervals (CrIs) are shown as error bars, centered around a point which is the mean of their lower and upper bounds. The diagonal is the identity line. Parameter recoveries with error bars spanning around the diagonal predict the true value within their CrI. Moreover, each panel shows the correlation between the true value and the point estimate as well as the normalized root mean squared error (NRMSE) of the CrI vs.\ the true value.}
\end{figure*}

Analogous to the inspection of the profiles log-likelihoods, we simulated data from the known model but generated 50 data sets, each with a unique combination of random parameter values within the bounds of the previously inspected intervals, effectively sampling from the prior distribution. Then, we fitted the model to each of the data sets, using uninformative uniform priors over the bounds shown in Figure~\ref{fig:likelihood-profiles}. Each fit is represented with one point per panel in Figure~\ref{fig:recovery}, showing 95\% credible intervals (CrIs) on the y-axis and the true parameter value on the x-axis. Ideally, CrIs would be narrow intervals spanning around the identity diagonal.

We can see that the 95\% CrIs almost always include the true value but are relatively wide, especially for the added parameters $F$, $d$, $\mu_{2}$, and $\mu_{3}$. Nevertheless, the agreement is generally good, as can be seen in the low normalized root mean square error or NRMSE values\footnote{The NRMSE is the mean root mean squared deviation from the true value across all samples of the posterior, normalized on the sample range.} and high correlations between true parameter values and CrI midpoints. This suggests that in general, true parameter values of simulated data sets can be recovered sufficiently well or at least with an acceptable level of uncertainty. As before, we note that parameter values, especially point estimates, should be interpreted with caution.

The reason for the high uncertainty for the new parameters is very similar to that for the profile log-likelihoods: Over the course of the entire fixation sequence, there are only very few retrieval events where these parameters could possibly have an effect on model behavior. Additionally, even when there is a retrieval, it is not guaranteed that it actually affects the activation of the currently fixated word, as the eyes may, for instance, already have continued past the retrieval trigger. Given these limitations, the recovery performance is surprisingly good, and the high correlations between true and recovered parameters appear very promising.

\subsubsection{Summary Statistics}
So far, we have demonstrated that SEAM, like SWIFT in its most current version \citep{Rabe2021}, can be successfully fitted to simulated data: The true parameter values are in the vicinity of profile log-likelihood peaks and are contained within parameter recovery CrIs. This means that if we assume the true underlying cognitive architecture to be similar to SEAM, we can reliably use fitted parameters (or their credible intervals) to make inferences about it. However, as the true underlying cognitive architecture is unknown, such checks are per se impossible on experimental data. Instead, we compare simulated and experimental behavior on the basis of relevant summary statistics. For this, as explained earlier, we first split the experimental data into a training and test subset, fitting the model to 70\% of the data of each participant and condition (training set), subsequently predicting eye trajectories for the other 30\% (test set).

\citet{Rabe2021} had previously noted that SWIFT, with the cross-validation method described above, is unable to make reliable predictions for regressive eye movements. However, given that SEAM now incorporates processes for cue-based memory encoding and retrieval, and given that memory retrieval processes are specifically hypothesized to trigger regressions by modulating the activation of retrieval candidates, in SEAM we should see an improvement in regression-related statistics such as incoming/outgoing regression probabilities, as well as regression path durations. These are also two important dependent measures in which effects were found in the experimental data set \citep[see {\em Experimental Study}, for a short summary; see][for details]{Mertz21}.

\begin{figure}
\caption{Spatial and Temporal Summary Statistics\label{fig:sumstats}}

\includegraphics{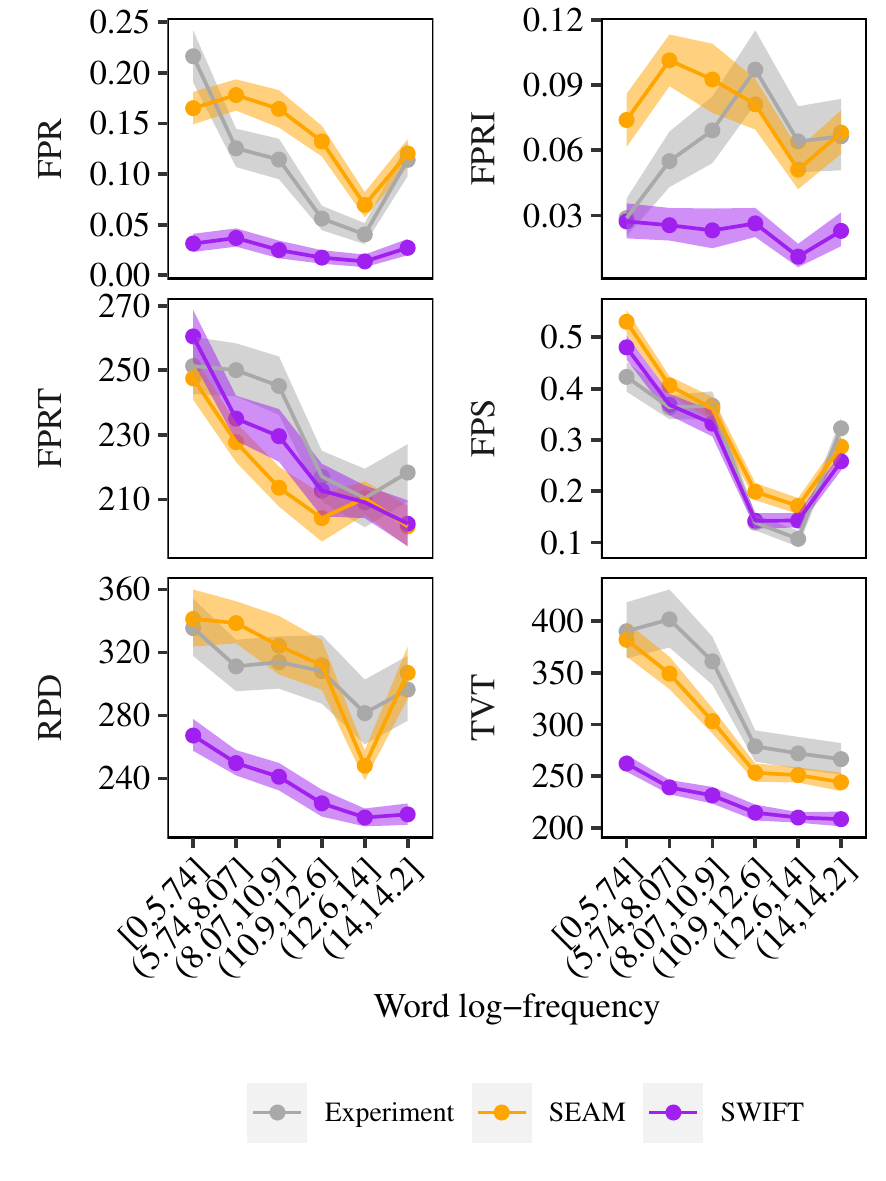}

\figurenote{Displayed are relevant regression-related fixation probabilities and durations as estimated grand means from a linear mixed-effects model. Ribbons are 95\% CrIs around the point estimate. Fixation targets (words) are grouped by log corpus frequency bins. \emph{FPR} = First-pass outgoing regression probability, \emph{FPRI} = First-pass incoming regression probability, \emph{FPS} = First-pass skipping probability, \emph{FPRT} = First-pass reading time (gaze duration), \emph{RPD} = Regression path duration (go-past time), \emph{TVT} = Total viewing/reading time. \mr{Words were not grouped into regions and all words of the sentences were considered.}}

\end{figure}

In Figure~\ref{fig:sumstats}, we show the comparison of summary statistics between experimental data and simulated data from the baseline SWIFT model (without memory retrieval) and SEAM (with memory retrieval). In all cases, SEAM predicts regression-related fixation probabilities and fixation durations more reliably than SWIFT. It is also noteworthy that not only the average across all word frequency bins but even word-frequency effects on summary statistics are reliably predicted.

\subsubsection{Experimental Effects of Memory Interference}

Arguably the most critical test for the SEAM architecture is to evaluate whether the model can predict differences in summary statistics between experimental conditions in the design of \citet{Mertz21}, which manipulates effects of memory retrieval on reading. 

\begin{table*}[!htbp]
\caption{Summary of Empirical vs.\ Model Estimates From SEAM and SWIFT of the Subjecthood and Animacy Effects on Regression Path Durations and First-Pass Regressions.}
\label{tab:results}
\begin{center}
\begin{tabular}{ccccccc}
\toprule
Region of interest & \multicolumn{2}{c}{Empirical estimates}        &    \multicolumn{2}{c}{SEAM}        &       \multicolumn{2}{c}{SWIFT} \\
 & subj  &         anim         &     subj  &  anim   &   subj  &   anim\\
\midrule
 & \multicolumn{6}{c}{Regression-path duration (ms)}\\
pre-critical  &  $[7, 77]$  &   $[-12, 60]$ & $[-5, 55]$ & $[-14,50]$ & $[-27, 12]$ & $[-33,6]$ \\
 critical verb        &  $[-6, 64]$   &   $[-16, 57]$ & $[-49, 58]$ & $[-37, 70]$ & $[-26, 16]$ & $[-25, 15]$\\
\midrule
 & \multicolumn{6}{c}{First-pass regressions (percentage)}\\
pre-critical  &  $[-2, 11]$ &   $[-4,9]$ & $[-3,6]$ & $[-3,7]$ & $[-2,1]$ & $[-1, 3]$ \\
 critical verb        &  $[2,15]$   &   $[-2,11]$ & $[0, 16]$ & $[-1, 16]$ & $[-3,2]$ & $[-4,1]$\\
\bottomrule
\end{tabular}
\end{center}
\tablenote{Shown are the 95\% credible intervals of the estimated effects from the data and from the two models. The empirical estimates are from the held-out data (30\% of the data). \emph{subj} = Effect of subjecthood, \emph{anim} = Effect of animacy.}
\end{table*}%

\begin{figure}
\caption{Posterior Distributions of Estimated Experimental Effects\label{fig:expeffects}}

\includegraphics{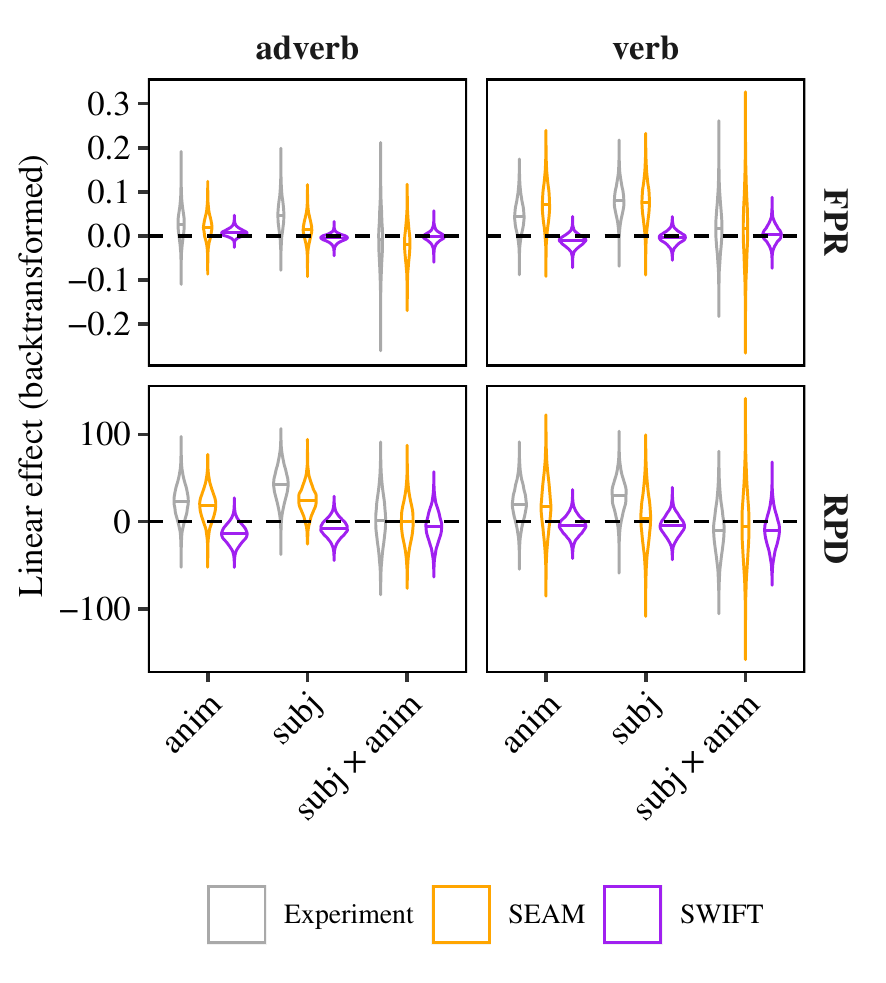}

\figurenote{Experimental effects on outgoing first-pass regression probabilities (FPR, top row) and first-pass regression path durations (RPD, bottom row), as found in the experimental data (gray), baseline SWIFT (purple), and SEAM (orange). Violin plots are posterior distributions of mixed-effects models.}

\end{figure}

\begin{figure}
\caption{Distribution of Absolute Prediction Errors for Estimated Experimental Effects\label{fig:compareeffects}}

\includegraphics{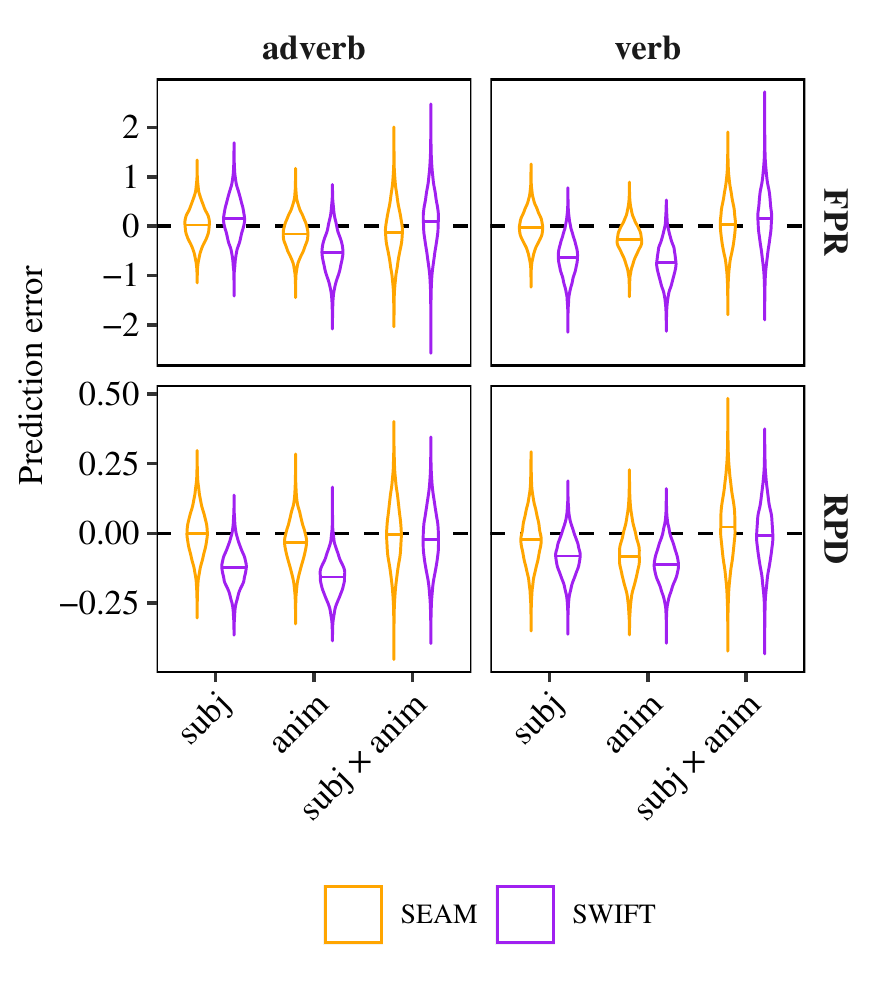}

\figurenote{Prediction errors of experimental effects on outgoing regression probabilities (top row) and regression-path durations (bottom row). Violin plots are paired differences of posterior distributions of baseline SWIFT vs.\ experimental data (purple), and SEAM vs experimental data (orange).}

\end{figure}

Based on a different experimental design, \citet{Rabe2021} were previously successful in demonstrating that SWIFT can be used to predict and explain differences in reading behavior when fitted to each participant and experimental condition separately. In our study presented here, however, we are only fitting one model at a time to each participant's data across all conditions, thereby considerably reducing the degrees of freedom. If the model is able to predict differences between experimental conditions, these do not originate from different parameter values for each condition but from the model dynamics, which are affected by the different feature specifications of the memory chunks across conditions. Therefore, capturing differences between conditions is a direct test of SEAM's added memory module. To illustrate the gain in empirical fit over baseline SWIFT, we also report predictions from SWIFT for reference. In SWIFT, no differences between experimental conditions are expected, because SWIFT has no parameters that could account for the processing cost of memory retrievals.

In order to evaluate the empirical fit of SEAM and baseline SWIFT, we conducted the same set of analyses for the observed experimental data and for the data predicted by SEAM and by SWIFT, after fitting each of the models to the training data sets. For both sets of data, we conducted a Bayesian mixed-effects regression for regression-path durations and outgoing regression probabilities as predicted by region and experimental condition (syntactic/semantic interference).

\label{r1-2.10}Table~\ref{tab:results}, and Figures~\ref{fig:expeffects} and \ref{fig:compareeffects} summarize the comparisons between the held-out empirical data and the predictions of SEAM and SWIFT. In order to interpret these comparisons, we compare SEAM and SWIFT against the empirical estimates from the held-out data using a region of practical equivalence (ROPE) approach  \citep{Freedman1984,spiegelhalter1994bayesian,kruschke2014doing} rather than formal model comparison methods such as k-fold cross validation, Bayes factors, or the like \citep[for tutorial introductions to these topics, see][]{NicenboimEtAlBayes2019}. The ROPE approach is a graphical model comparison method that involves comparing model predictions against observed estimates from data; overlap in the posterior distribution of estimates provides an informal basis for deciding whether a model approximately matches observed estimates. In this approach, there is no notion of {\em statistical significance}; rather, the focus is on whether the model predictions are approximately consistent with the data. One important reason for taking this informal model comparison approach is the fact that the held-out data are relatively sparse. For this reason, the present evaluation should be seen rather as a proof-of-concept rather than a comprehensive evaluation. Such an evaluation would require significant amounts of benchmark data \citep[for examples of such extensive evaluations, see][]{EngelmannJaegerVasishth2019,Yadavetal2022,NicenboimPreactivation2019} and must be left for future work.

\begin{figure}
\caption{\mr{Effect of Subjecthood on the Contributions of First-pass and Second-pass reading times to Regression Path Durations}\label{fig:rpd-contributions}}

\includegraphics{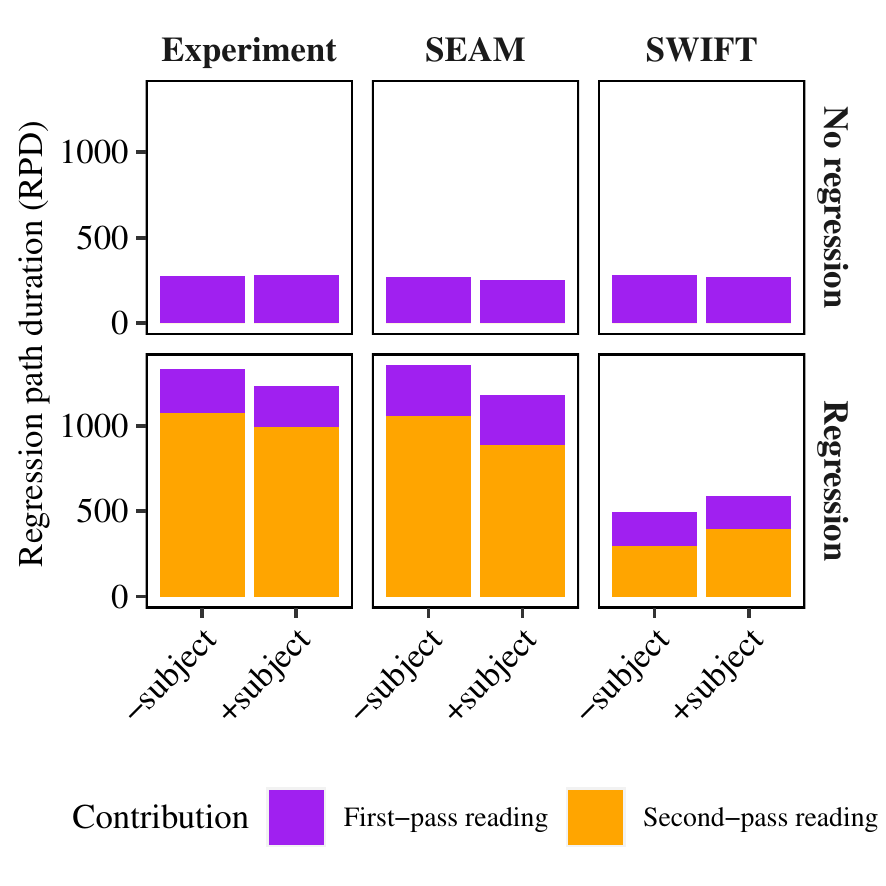}

\figurenote{\mr{Columns are sums of first-pass reading time on the launch site and rereading times on refixated previous regions until the (simulated) eye leaves to the right of the initial launch site.}}

\end{figure}

Table~\ref{tab:results}, and Figures~\ref{fig:expeffects} and \ref{fig:compareeffects} show that the predictions for the experimental effects of animacy (semantic interference) and subjecthood (syntactic interference) in the experimental data are generally more in agreement with SEAM than with SWIFT: the violin plots in Figure~\ref{fig:expeffects} from SEAM have a better overlap than the observed data than the predictions from SWIFT. This is true in both the pre-critical and critical regions, in both the first-pass regression and regression path duration measures. One exception is the subjecthood effect at the critical verb (see the bottom right \mr{\label{r1-1.12}panel} in Figure~\ref{fig:expeffects}); SEAM predicts essentially no effect of subjecthood, just like SWIFT. \mr{This is mainly because the regression paths predicted by SEAM are somewhat too short, i.e.\ return too early, in the +subject conditions (see Figure~\ref{fig:rpd-contributions}).} We return to this in the Discussion section.

Given that SWIFT does not have any mechanism that accounts for cue-based memory retrieval, it is expected that the model predicts no effects of memory interference. 
Notice that the violin plots for the data as well as the SEAM and SWIFT predictions shown in  Figures~\ref{fig:expeffects} and \ref{fig:compareeffects}  are relatively wide; this is due to the fact that only  30\% of the test portion of experimental data (the held-out data) are compared to the model predictions.

\begin{figure*}
\caption{Effects of Memory Interference on Experimental and Simulated Regression Probabilities}

\label{fig:regprobeffects}

\includegraphics{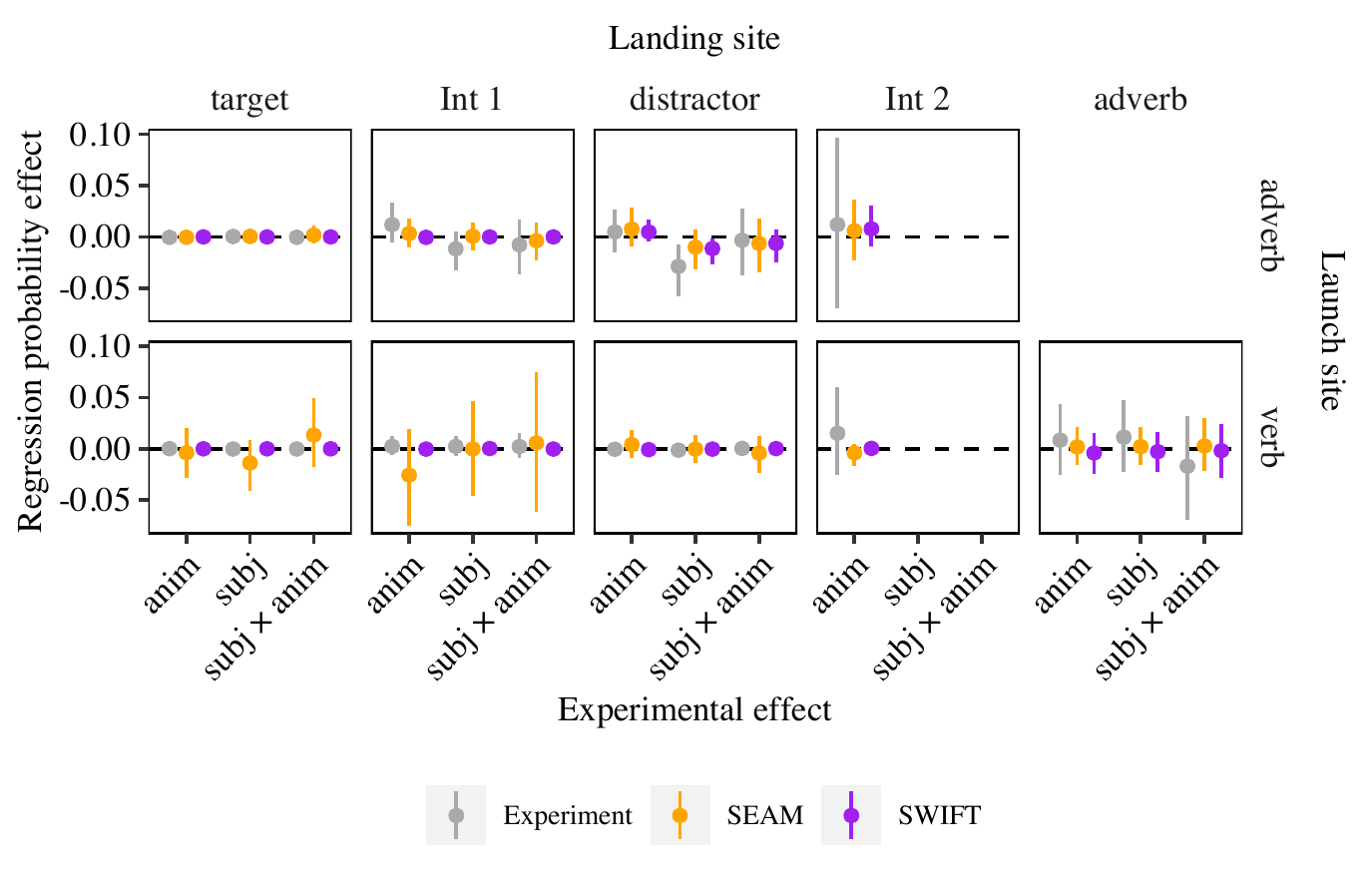}

\figurenote{\mr{Effects are estimates and 95\% CrIs from logistic regressions. Each panel shows the effect on the proportion of trials with a critical regression given the launch site (rows) and landing site (columns). Estimates have been backtransformed to the linear scale. Regions Int~1 and Int~2 are intervening regions between target and distractor, and between distractor and adverb, respectively.} \mrtwo{Note that there is no Int2 in conditions a and b, so there is no subjecthood effect or interaction in that region.}}

\end{figure*}

\begin{figure*}
\caption{\mr{Conditional Means of Experimental and Simulated Regression Probabilities}\label{fig:regprobmeans}}

\includegraphics{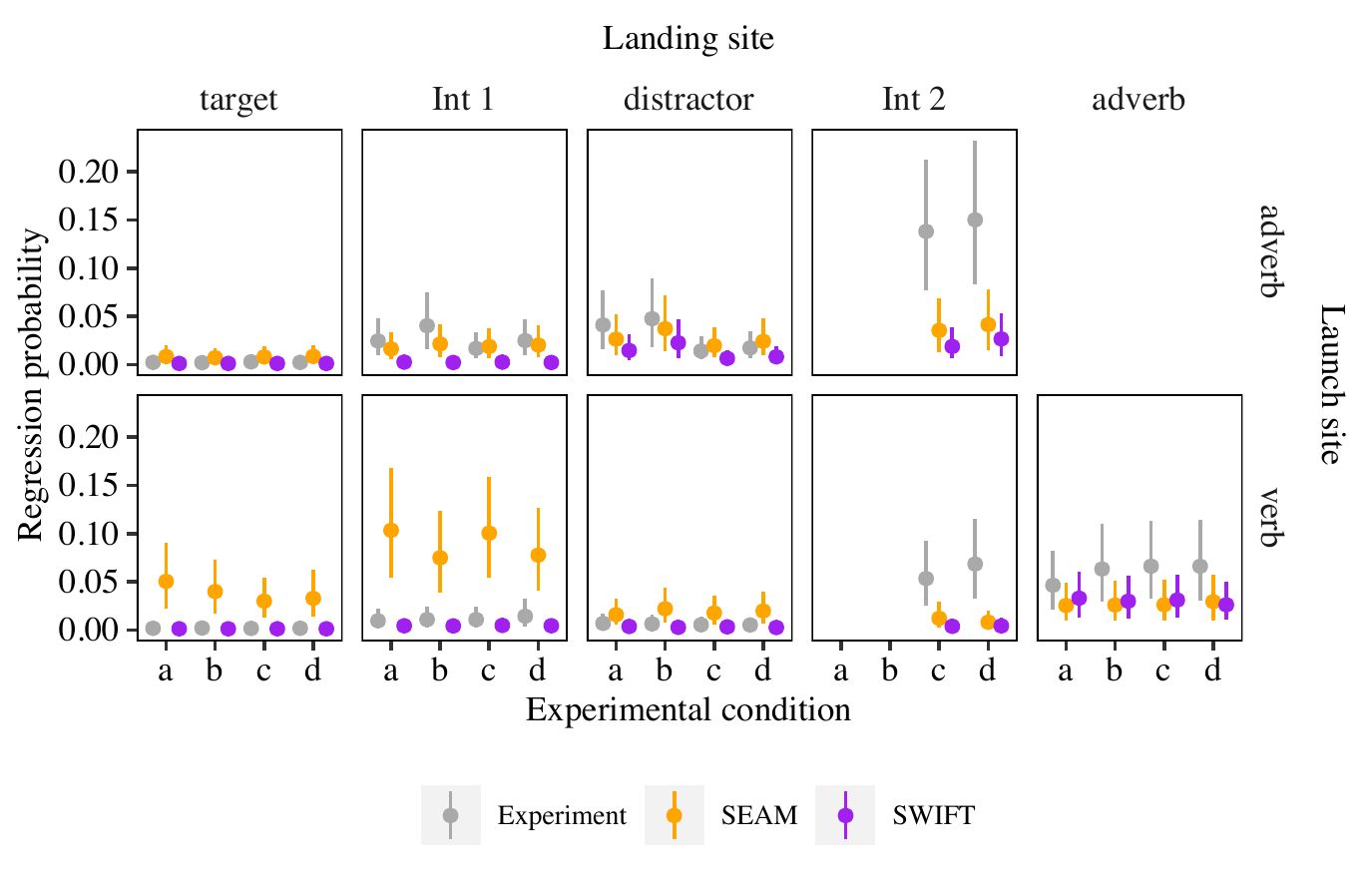}

\figurenote{\mr{Shown are estimates and 95\% CrIs from nested logistic regressions. Each panel shows the mean proportion of trials with a critical regression given the launch site (rows) and landing site (columns). Estimates have been backtransformed to the linear scale. Regions Int~1 and Int~2 are intervening regions between target and distractor, and between distractor and adverb, respectively. Conditions a--d refer to the four conditions --subj/--anim, --subj/+anim, +subj/--anim, and +subj/+anim, respectively, as shown in Example~\ref{ex2}.} \mrtwo{Note that there is no Int2 in conditions a and b.}}

\end{figure*}

\begin{figure*}
\caption{\mr{Conditional Means of Experimental and Simulated Regression Durations}\label{fig:regdurmeans}}

\includegraphics{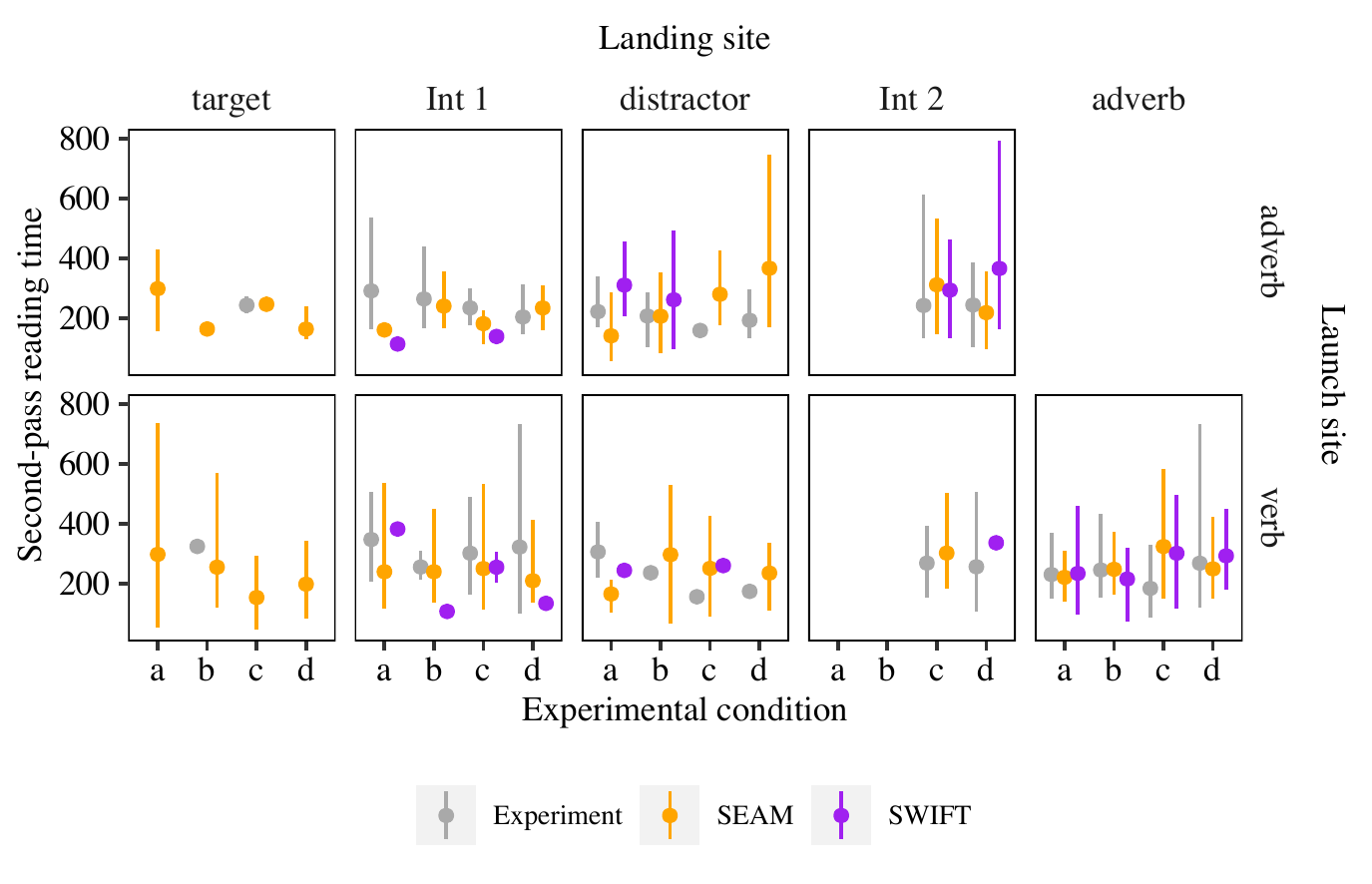}

\figurenote{\mr{Shown are means and 95\% quantiles of the raw data. Each panel shows the mean second-pass reading time (regression duration) following a critical regression given the launch site (rows) and landing site (columns). Regions Int~1 and Int~2 are intervening regions between target and distractor, and between distractor and adverb, respectively. Conditions a--d refer to the four conditions --subj/--anim, --subj/+anim, +subj/--anim, and +subj/+anim, respectively, as shown in Example~\ref{ex2}.} \mrtwo{Note that for some cells, there were not enough data to visualize due to the low regression probabilities (see Figure~\ref{fig:regprobmeans}).}}

\end{figure*}

\mr{\label{r1-reg-analysis}A main motivation of SEAM was to develop a model in which low-level psychological and high-level linguistic processes interact. \mrtwo{\label{r2-reg-analysis}The integration of the LV05-based memory module is expected to influence eye movements, particularly in situations where demanding dependency resolution is required. This effect may be especially pronounced when there is a high level of ambiguity between the correct dependents and distractors.} Even though we already know that the \citet{Mertz21} data do not provide unequivocal evidence in support of \mrtwo{an influence of memory retrieval on eye movements}, we can look at the distribution of regressions across trials conditional on launch and landing sites in order to investigate where regressions from the (pre-)critical region tend to land in the experimental data and in the simulations. Figure~\ref{fig:regprobmeans} and Appendix~\ref{fig:regprobeffects} show that regressions in general have a tendency to land on the preceding word.\footnote{\label{fn:r2-int2}\mr{Note that in \mrtwo{the --subj} conditions a and b in Figure~\ref{fig:regprobmeans}, the distractor immediately precedes the pre-critical adverb, while conditions c and d have an intervening region between the distractor and pre-critical adverb.} \mrtwo{There is no intervening region (Int2) between distractor and adverb in the --subj conditions a and b.}} In these cases, SEAM is in better agreement with the experimental data than SWIFT. For regressions launched from the verb, however, SEAM currently predicts too many regressions on average, although the experimental effects (i.e., differences between conditions, see Appendix~\ref{fig:regprobeffects}) are still in agreement with the experimental data. As there are generally very few regressions, both in the experimental and in the simulated data, analysis of regression durations is problematic but Figure~\ref{fig:regdurmeans} shows that they are also generally in good agreement with each other.}

\begin{figure*}
\caption{Effects of Experimental Condition on SEAM Word Activations at Encoding of the Critical Verb\label{fig:activationeffects}}

\includegraphics{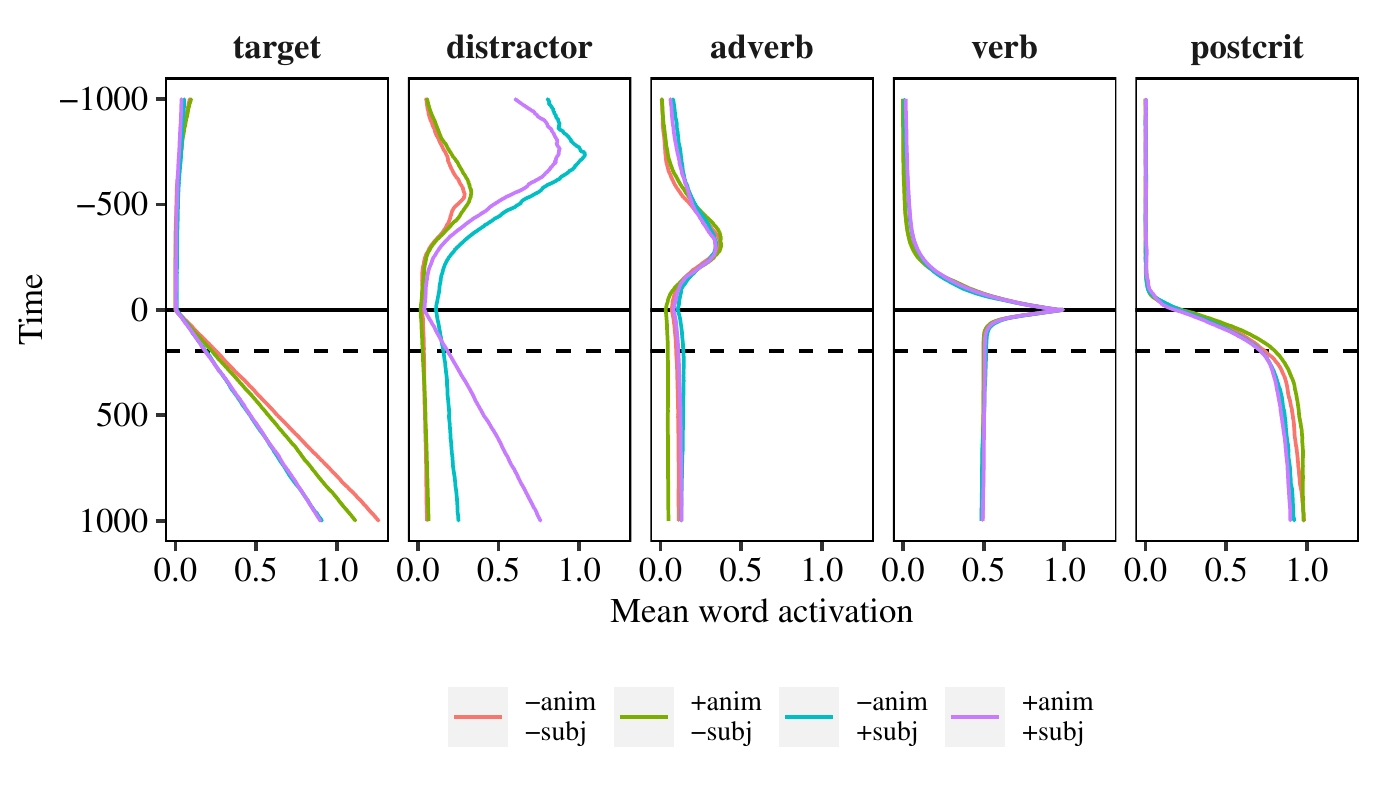}

\figurenote{SEAM word activation, \mrtwo{$a_j(t)$}, of the target, distractor, pre-critical adverb, critical verb, and post-critical region of a sentence grouped by experimental condition. Activations are averaged across 500 independent simulations of the same item in all four conditions. For each simulation, $t=0$ \mrtwo{(solid horizontal line)} is adjusted to the time of the start of post-lexical processing of the critical verb, that is, the start of the retrieval. \mrtwo{The average time for the programming of the upcoming saccade, i.e.\ the end of the non-labile saccade programming stage, is marked by the dashed horizontal line.}}

\end{figure*}

As SEAM and SWIFT are nested models,\footnote{SWIFT is nested within the more complex SEAM, and the effects of the memory retrieval submodule on the word activations can be completely turned off by setting $F=0$ and $\mu_3=0$ because that immediately ends any started retrieval and does not reactivate any previously encoded words.} the fact that SEAM but not SWIFT can predict different summary statistics is a first indicator that the differences in predictive power between the models may be due to the added memory retrieval submodule. To verify this and to attempt an explanation of the differences in observed behavior, we look at the differences in the internal model dynamics under the different experimental conditions.

In particular, we can examine the word activation field, which is the main driver for target selection probabilities in SEAM and SWIFT (Equation~\ref{eq:selprob}), including regressive saccades. In Figure~\ref{fig:activationeffects}, we show word activations in SEAM, averaged across 500 independent simulations, using the mean estimated model parameters across all model fits. Before averaging across simulations, all word activations are centered on the temporal dimension so that $t=0$ is the time when the activation of the critical verb reaches its maximum, that is, when post-lexical processing of the critical verb starts and triggers the memory retrieval.

First, it is important to note that the activations of the critical verb, when normalized in time, do not vary substantially between experimental conditions. Although some conditions seem to have a slower decrease than others, overall the curves are very similar in all conditions. When the retrieval starts at $t=0$, retrieval candidates are reactivated, with their memory activation $A_{j,n}'\left(t\right)$ modulating the transition rate $w_j'\left(t\right)$ (see Equation~\ref{eq:wword-seam}) of word/memory chunk $j$. While the activation for the target word seems to be very similar over time between conditions as well, there is some variability in the time course of the activations of the distractor noun and of the adverb around the retrieval.

Regarding the adverb, the main reason it is reactivated during retrieval is that it has the highest \mr{base-level activation} $B(t)$, as it was most recently encoded/accessed in memory before the retrieval started. The later processing of object noun distractors also attenuates the processing that the adverb receives, which leads to weaker reactivation of the adverb during the retrieval.

We can also observe that the distractor word activations prior to the retrieval peak earlier for the two conditions where the distractor is a subject noun, that is, in the conditions where there is syntactic interference. This effect is not related to the retrieval at the critical verb (which has not started at this time), but is due to the distractor appearing earlier in the sentence when it is a syntactic subject. \label{r2-4.2a}Interestingly, the distractor noun only significantly peaks during the retrieval in the +animate/+subject condition, that is, when both features match the retrieval cues. The distractor thus only attracts regressions when both the animacy and subjecthood features match, i.e., when there is both syntactic and semantic interference. Despite this difference in word activations, there is no significant difference between the proportions of observable targeted regressions from the critical verb to the distractor noun between any of the experimental conditions. This is true for the experimental data as well as for the data simulated by SEAM and SWIFT, as shown in \mr{Figure~\ref{fig:regprobmeans} and Appendix~\ref{fig:regprobeffects}}.

\label{r2-4.2b}As the estimates show, there is no indication in the experimental data that the distractor is regressed to more often in the +animate/+subject condition. The distractor's activation pattern in Figure \ref{fig:activationeffects} is simply a consequence of the hard-coded assumption in LV05 that it has the highest feature match in this condition. Interestingly, however, the predicted data from SEAM do not show an increase in incoming regressions \mr{to the distractor} either. An increase in word activation thus does not necessarily translate into a change in observed eye movements. The lack of a direct effect on distractor refixations is likely due to oculomotor error and because \mrtwo{the target word and} upcoming words have even higher activations \mrtwo{relative to the} distractor \mrtwo{at the time the saccade target is programmed (horizontal dashed line in Figure~\ref{fig:activationeffects})}.

\mr{Based on results from preliminary post-hoc analyses, the overestimation of the average regression probability from the critical verb to preceding regions (see Figure~\ref{fig:regprobmeans}) is also probably due to the hard-coded retrieval schedules. Even though the times of memory encoding, and therefore the base-level activation, are stochastic and completely governed by the eye's trajectory, the feature match is deterministic. Future work could investigate alternative links between memory activation $A_{j,n}'\left(t\right)$ and word activation $a_j(t)$ or transition rates $w_j'(t)$, as they currently implement a very strong linking hypothesis.}

\mrtwo{\label{r2-4.1}Furthermore, the set of fitted SWIFT parameters is relatively small. For example, we did not fit parameters that modulate oculomotor execution, corpus frequency effects, word difficulty, predictability etc. Previous studies had already shown that SWIFT has a general tendency to predict too low regression probabilities \citep{Rabe2021}. This is also apparent in the relatively low simulated regression probabilities in SWIFT and SEAM from adverb to preceding intervening region or from verb to adverb (see Figure~\ref{fig:regprobmeans}). It would be worthwhile to fit more parameters at once, also including previously established SWIFT parameters, to allow the MCMC sampler to find a more plausible equilibrium between the different mechanisms in the model that can cause regressive eye movements. It is very likely that considering more free parameters will also improve the fit with regard to predicted regression locations. In this particular manuscript, however, we specifically focused on selected parameters demonstrating that such an integrated model is possible to be implemented within a Bayesian inference framework.}

\mr{In this context, we also note that -- as far as we are aware -- the only study that has previously looked at word-level rereading as a function of similarity-based interference is \citet{Lee07}. The authors report longer rereading times for a sentence-initial region containing both the retrieval target and the distractor in their high-interference conditions, but the Korean sentences used in their study were relatively short compared to those used by \citet{Mertz21}. In future work, shorter sentences should be a fruitful testing ground for SEAM. If SEAM generates more linguistically mediated targeted regressions in shorter sentences, this would be in line with human data \citep{Inhoff2005}.}

\subsubsection{Summary}
We showed that both SEAM and SWIFT can be fitted to the \citet{Mertz21} experimental data set. In contrast to SWIFT, however, SEAM's predictions are in good agreement with the overall and by-frequency regression probabilities and regression-path durations. SEAM shows the more specific memory interference effects, that is, differences in regression probabilities and regression-path durations due to differences in the animacy and subjecthood of a distractor noun. 

Given that the compared models SEAM and SWIFT only differ in the supplemental cue-based memory retrieval processes contributed by the LV05 component, we can attribute the better performance of SEAM in these metrics to LV05 principles with the four additional  parameters that were fit to the training data from \citet{Mertz21} ($F$, $d$, $\mu_2$, and $\mu_3$). It is also noteworthy that these parameters were estimated based on a restricted training data set for each participant, and that the model can make reasonable predictions on the held-out test data for all experimental conditions with a single model fit for each participant.

Furthermore, even though the models are compared to each other and to the experimental data using summary statistics and predicted experimental effects, neither SWIFT nor SEAM was directly optimized to reproduce these measures. Instead, both models were fitted directly to the raw, unbiased fixation sequences of each participant. Therefore, the models can make reasonably accurate predictions for summary statistics and experimental effects although they are not specifically fitted to them.

\subsection{Discussion}

We showed that adding a memory interference mechanism in the SWIFT architecture---resulting in the SEAM model---allows us to bring together eye-movement control theory and a psycholinguistic  account of dependency completion. We demonstrated that that the key regressive eye-movement related patterns in an experimental psycholinguistic data set can be accounted for by the SEAM architecture. Specifically, we showed that first-pass regressions and regression path duration patterns that occur due to the interference manipulation in the \citet{Mertz21} data can be accounted for by SEAM, but not by SWIFT; in SEAM, as in the data, both syntactic and semantic interference have an impact on the two dependent measures at the pre-critical region and the critical verb. 

The main results of our simulations are summarized in Table~\ref{tab:results} and Figures~\ref{fig:expeffects}, \ref{fig:compareeffects}, and \ref{fig:activationeffects}.
There were three interesting patterns in the SEAM fit that deserve discussion. First, as shown in Figure~\ref{fig:expeffects}, at the critical verb, regression path durations from SEAM show essentially no effect of subjecthood; this is surprising because the data do show such an effect. At the same time, in SEAM, first-pass regressions at the verb show a clear subjecthood effect. This \mr{is because} even though regressions were triggered at the verb, \mr{\label{r1-2.8}which should itself increase the mean RPD, regression paths predicted by SEAM return too early in the +subject conditions, thereby masking the effect on RPD\footnote{\label{fn:maskedeffect}As the regression path duration is the sum of gaze durations on the current word and on all preceding regions until the (simulated) eye leaves to the right of the current word, an effect in regression path durations could be due to (a) an effect on gaze durations on preceding regions, (b) an effect on gaze duration on the current word, (c) a combination of both. Likewise, a null effect could be a masked effect of gaze durations on the \mr{launch site} vs.\ gaze durations on preceding regions.}.}

The second interesting pattern relates to the effects observed at the pre-critical adverb region \mr{\label{r1-2.11}(\emph{the attorney whose secretary had forgotten […] \underline{frequently} complained}, see Example~\ref{ex2})}. Recall that in the original LV05 model, sentences are processed in strictly serial order. Effects of similarity-based interference at the pre-critical adverb are thus unexpected under this model: Given the assumption that the verb is the retrieval trigger, there should be no retrieval-related effects before it is read. Nevertheless, \citet{Mertz21} did observe interference effects at the pre-critical adverb  \citep[others have found similar patterns in the pre-critical region; see ][]{VanDyke2007, Lago2021}. Mertzen and colleagues discuss several possible reasons for these effects: Differential processing spillover from previous regions due to differences in sentence complexity between conditions, lingering memory interference during encoding of the noun phrases, and predictive processing of the verb. A final important possibility considered by \citet{Mertz21} is parafoveal preview of the verb while the adverb is being processed, so that the verb can trigger the retrieval prior to being fixated. Our SEAM simulations are partly consistent with this last account: In 25\% of our simulations, the verb reaches the retrieval stage while the adverb is being fixated. However, there is also processing spillover in the form of residual word activation in SEAM. Especially in the +subject conditions, where there is an additional retrieval in the embedded sentence at \textit{was important}, and the activation of the retrieval target may not have fully decayed when the adverb is read, leading to more regressions. Based solely on the \citet{Mertz21} data and the small sample size of the held-out data, it is difficult to quantify the relative contributions of preview and spillover, and we leave this issue to future research. Nevertheless, SEAM provides a promising starting pointing for tackling possible pre-critical retrieval effects.

A third noteworthy pattern occurs in Figure~\ref{fig:activationeffects}; the +subject/+animate condition causes a large increase in the distractor's word activation after the critical verb is encoded. This suggests that the probability of the distractor to attract regressions should be much higher in that condition than the sum of the +subject/--animate and --subject/+animate conditions. Even though the combination of the two retrieval cues is additive at the level of the LV05 memory activation (see Equation \ref{eq:memory-activation}), the exponential transformation of $A(t)$ in Equation~\eqref{eq:wword1} significantly amplifies it. Nevertheless, the superadditive effect on the distractor's activation when it matches both retrieval cues does not generate any detectable overadditive effects in the analyzed regression-related dependent measures (regression path duration and first-pass regressions). As discussed in the previous section, the spike in activation does not necessarily translate into observed regressions, partly because the large distance between the verb and the distractor amplifies the influence of oculomotor error. With less complex sentences, it is thus possible that SEAM would show effects on the observed regression probabilities.

\section{General Discussion}

From the very beginning of eye-movement research in reading, a dominant idea has been that the eye and mind are tightly coupled \citep[e.g.,][]{Just1980}. After psycholinguists started looking at fixation patterns in reading as a function of language comprehension difficulty, an important idea that was expressed in a now-classic paper by \citet{Frazier1982} was the selective reanalysis hypothesis: this was the idea that increased comprehension difficulty (e.g., due to garden-pathing) leads to targeted regressions to a preceding region that caused the processing difficulty.  Although the strongest version of selective reanalysis, \mr{and thus of the eye-mind assumption}, is difficult to uphold given subsequent investigations \citep[e.g.,][]{Mitchell08,malsburgvasishthsub10}, it is nevertheless well-established that increased regressions are triggered when language processing difficulty occurs \citep[e.g.,][]{cliftonjr:emr}, and that rereading can aid comprehension \citep{Schotter14}. \mr{\label{r1-0.2a}We assume that the mixed evidence in the psycholinguistic literature regarding selective rereading (see \cite{Paape22} for a review) may be the result of a more indirect linkage between higher-level sentence processing and saccade targeting: In our model, retrieval events during dependency completion affect the activation values of previous words in the sentence. Words with higher activation will tend to attract saccades, but due to the inherent stochasticity of the eye-movement control system and oculomotor error, subtle linguistic manipulations do not necessarily engender measurable effects at typical sample sizes.}

Most of the psycholinguistic work carried out on reading until now has side-stepped the underlying complex latent processes involved in reading, and instead focused only on key events involved in linguistic dependency completion. Abstracting away from these underlying latent reading processes has had many advantages, a major one being that it allows us to focus exclusively on the psycholinguistically interesting aspects of processing at the level of the sentence representation. On the other hand, the simplification comes at a cost, because interactions between constraints on eye-movement control and language comprehension end up being ignored. 

Interestingly, cognitive psychology has gone in a completely different direction than psycholinguistics: there, the focus has been on spelling out detailed process models of eye-movement control that rely primarily on relatively low-level drivers of eye movements, such as frequency and word length. Models of eye-movement control such as E-Z Reader \citep{Reichle1998} and SWIFT \citep{Engbert2005} have shown excellent performance in explaining benchmark data in reading,  without modeling the higher-level cognitive processes such as linguistic dependency completion in any great detail. 

One major gap in the literature is that these two threads---psycholinguistic explanations of reading difficulty versus cognitive psychology models of reading---have only rarely been considered to be joint actors in explaining key effects observed in experimental data from psycholinguistics. Our paper makes an attempt to fill this gap: using data from a classic similarity-based interference design, we demonstrate one way in which an eye-movement control model, SWIFT, can be extended to include dependency completion processes. We show that such an extended model (SEAM) can produce regressive eye movements triggered by retrieval that occurs during linguistic dependency completion.  Developing such models is the only way to unpack the latent processes involved in reading and to investigate how \emph{low}- and \emph{high}-levels of cognitive processes interface dynamically.  To our knowledge, SEAM is the only model to date that extends a complete model of eye-movement control with a detailed model of linguistic dependency completion, using data from a planned experiment in psycholinguistics and rigorous statistical inference.

Apart from using SWIFT as the eye-movement module, SEAM differs in important ways from previous integrative models of eye movement control and higher-level sentence processing. For instance, Über-Reader \citep{Reichle2020}, whose eye movement module is highly similar to that of E-Z Reader \citep{Reichle1998}, has a parsing module that builds syntactic structure, but each parsing step is assumed to take the same amount of time. In SEAM, by contrast, completing syntactic dependencies takes a variable amount of time that is determined by the LV05 Equations (which originally come from ACT-R). Furthermore, regressive saccades are not captured by Über-Reader, but are modeled dynamically in SEAM.

Another integrative model proposed by \citet{Dotlacil2021}, whose eye movement module is also based on \mr{\label{r1-1.14}E-Z Reader}, makes use of ACT-R Equations, but in a different way from SEAM: In \citeauthor{Dotlacil2021}'s model, the latency with which a given dependent word is integrated into the sentence's syntactic representation depends on the retrieval time for the dependent words and additionally on the retrieval time for the relevant parsing rule from declarative memory. SEAM does not assume retrieval of parsing rules, which are assumed to be represented as procedural knowledge, as in the LV05 model. Another salient difference between the models is that regressions in \citeauthor{Dotlacil2021}'s (\citeyear{Dotlacil2021}) model are only triggered when parsing failure occurs, while regressions in SEAM are driven by the dynamic target selection processes taken over from SWIFT. As a final comparison, the model of \citet{Engelmann2013} and \citet{VasishthEngelmann2020} combines an LV05 sentence processing module with eye movement control based on EMMA \citep{Salvucci2001}, but also does not provide a detailed model of saccade targeting, unlike SEAM.

There are of course several limitations to the present work.  First and foremost, the current implementation of SEAM and its evaluation are only a proof-of-concept. Because of the absence of large-scale data sets with psycholinguistically interesting manipulations, it is difficult to present a comprehensive evaluation of the proposed SEAM architecture. However, such an investigation is in principle possible to carry out, given (i) the progress on Bayesian inference for process-based models and (ii) the fact that more and more researchers are releasing data and code associated with their published papers. We expect that in future work, more comprehensive evaluations of architectures like ours can be carried out, using large-scale data from a broad range of phenomena in psycholinguistics. At a minimum, such an investigation would need to include cross-linguistic data from garden-path sentences of different types \citep[e.g.,][]{frazier79}, predictability manipulations   \citep[e.g.,][]{levy08}, the full spectrum of similarity-based interference effects \citep[e.g.,][]{JaegerEngelmannVasishth2017}, underspecification effects \citep[e.g.,][]{swets2008underspecification}, etc. This would be a sizable project, but one which would significantly advance our understanding of how eye-movement control and parsing interface during reading.

A second limitation is that, due to the computational complexity of investigating such a detailed model of reading, formal model comparison between the baseline SWIFT model and the SEAM model is difficult to carry out. We avoided overfitting the models to data by separating the empirical data into a training set and a held-out set, and evaluating the model fit only on the held-out set. This is already a significant advance over conventional approaches to model evaluation; in both cognitive psychology and psycholinguistics, it is common to evaluate a model on the same data that it is trained on. In principle, it is possible to go even further than we did in this paper, and to evaluate predictive performance by using k-fold cross-validation. This would  involve creating $k$ (usually, in machine learning, $k=10$) subsets of the data to train on, and then use the $k$ held-out data sets for evaluation; this would allow us to compute a quantitative measure of average fit, such as expected log pointwise density \citep[e.g.,][]{gelman2014understanding}. We did not carry out such a quantitative evaluation because it would have been computationally extremely costly. For example, just the pure SWIFT model discussed in \citet{Rabe2021} required a high-performance computing environment, and the total computing time was approximately 10,000 core hours, amounting to 3.5 hours run time on 72 independent parallel nodes with 40 cores per node. Our goal in the present work was to get as close as possible to the underlying processes involved in reading, but obviously this comes with an unavoidable computational cost. 
 
\section{Conclusion}
 
We present an integrated model of eye-movement control and linguistic dependency completion while reading. The model, called SEAM, is an integration of the SWIFT model of eye-movement control and the Lewis-Vasishth model of sentence processing. SEAM is evaluated using experimental data from a similarity-based interference experiment. We show that the SEAM model can account for empirically observed regressive eye movements; in the model, regressive eye movements are shown to be triggered by retrieval processes that result from higher-level dependency completion during sentence parsing. To  our knowledge, this is the first demonstration of how eye-movement control and sentence comprehension processes can interact in explaining data from a psycholinguistically controlled experiment.

\section{Acknowledgements}

This work was funded by the \foreignlanguage{ngerman}{Deutsche Forschungsgemeinschaft} (DFG, German Research Foundation), projects~317633480 (SFB~1287 \emph{Limits of Variability in Language}) and 318763901 (SFB~1294 \emph{Data Assimilation}). We also acknowledge support by \foreignlanguage{ngerman}{Norddeutscher Verbund für Hoch- und Höchstleistungsrechnen} (HLRN, project no.~bbx00001) for providing high-performance computing resources.

\printbibliography

\clearpage

\begin{longtable}[l]{lr@{\extracolsep{0pt}.}ll}
\caption{SEAM Model Parameters\label{app:parameters}}\\%
\toprule 
\addlinespace
Parameter & \multicolumn{2}{c}{Default} & Description\tabularnewline
\midrule
\endhead
$F$ & 0&14 & Retrieval latency scaling factor\tabularnewline
$S_\mathrm{max}$ & 1&5 & Maximum activation strength\tabularnewline
$d$ & 0&5 & Exponential decay of memory activation\tabularnewline
$p$ & --1&5 & Additional penalty for word-category mismatch\tabularnewline
$\mu_{1}$ & 0&0 & Fixed production time (50~ms in original LV05 model) \tabularnewline
$\mu_{2}$ & 0&5 & Relative minimum activation of retrieval trigger during retrieval \tabularnewline
$\mu_{3}$ & 0&2 & Relative activation threshold for retrieval candidates \tabularnewline
\midrule
$\delta_{0}$ & 7&23 & Non-dynamical (fixed) processing span width (in letter spaces)\tabularnewline
$\delta_{1}$ & 1&0 & Dynamical processing span width (in letter spaces)\tabularnewline
asym & 1&0 & Relative width of the processing span to the left of the fixation location\tabularnewline
$\eta$ & 0&5 & Word-length exponent\tabularnewline
$\alpha$ & 1&50 & Baseline word difficulty\tabularnewline
$\beta$ & 0&5 & Word-frequency effect on word difficulty\tabularnewline
$\gamma$ & 1&0 & Target selection exponent\tabularnewline
minact & --5&0 & Minimum activation threshold of words for target selection\tabularnewline
$\theta$ & 0&0 & Effect of predictability on processing speed\tabularnewline
$t_{\mathrm{sac}0}$ & 1&0 & Relative duration of the first fixation of the sequence\tabularnewline
$t_{\mathrm{sac}}$ & 2&2 & Mean saccade interval (fixation duration)\tabularnewline
$h$ & 0&64 & Foveal inhibition factor\tabularnewline
$h_{1}$ & 0&0 & Parafovial inhibition factor\tabularnewline
ppf & 0&0 & Inhibition from words to the left of the fixation location\tabularnewline
$\iota$ & 1&0 & Transfer across saccades (activation loss during saccade)\tabularnewline
$M$ & 1&25 & Relative fixation duration of misplaced fixations\tabularnewline
$R$ & 0&8 & Relative fixation duration of well-placed refixations\tabularnewline
$\kappa_{0}$ & 0&0 & Non-labile latency dependence on target distance (factor)\tabularnewline
$\kappa_{1}$ & 0&0 & Non-labile latency dependence on target distance (exponent)\tabularnewline
proc & 1&0 & Relative processing speed for postlexical processing\tabularnewline
decay & 0&07 & Global decay of word activations during postlexical processing\tabularnewline
$\tau_{\mathrm{l}}$ & 1&2 & Mean duration of the labile saccade program\tabularnewline
$\tau_{\mathrm{n}}$ & 0&8 & Mean duration of the non-labile saccade program\tabularnewline
$\tau_{\mathrm{x}}$ & 0&2 & Mean duration of saccade execution\tabularnewline
aord & \multicolumn{2}{c}{30} & Order of random walks for word activation\tabularnewline
cord & \multicolumn{2}{c}{15} & Order of random walks for global saccade program\tabularnewline
lord & \multicolumn{2}{c}{12} & Order of random walks for labile saccade program\tabularnewline
nord & \multicolumn{2}{c}{10} & Order of random walks for non-labile saccade program\tabularnewline
xord & \multicolumn{2}{c}{20} & Order of random walks for saccade execution\tabularnewline
ocshift & 0&0 & Fixed oculomotor shift parameter\tabularnewline
$\mathrm{omn}_{1}^{\mathrm{(FS)}}$ & 0&80 & Oculomotor noise intercept parameter for forward saccades\tabularnewline
$\mathrm{omn}_{2}^{\mathrm{(FS)}}$ & 0&03 & Oculomotor noise slope parameter for forward saccades\tabularnewline
$\mathrm{omn}_{1}^{\mathrm{(SK)}}$ & 0&80 & Oculomotor noise intercept parameter for skipping saccades\tabularnewline
$\mathrm{omn}_{2}^{\mathrm{(SK)}}$ & 0&14 & Oculomotor noise slope parameter for skipping saccades\tabularnewline
$\mathrm{omn}_{1}^{\mathrm{(FRF)}}$ & 0&80 & Oculomotor noise intercept parameter for forward refixations\tabularnewline
$\mathrm{omn}_{2}^{\mathrm{(FRF)}}$ & 0&03 & Oculomotor noise slope parameter for forward refixations\tabularnewline
$\mathrm{omn}_{1}^{\mathrm{(BRF)}}$ & 0&80 & Oculomotor noise intercept parameter for backward refixations\tabularnewline
$\mathrm{omn}_{2}^{\mathrm{(BRF)}}$ & 0&03 & Oculomotor noise slope parameter for backward refixations\tabularnewline
$\mathrm{omn}_{1}^{\mathrm{(RG)}}$ & 0&80 & Oculomotor noise intercept parameter for regressions\tabularnewline
$\mathrm{omn}_{2}^{\mathrm{(RG)}}$ & 0&03 & Oculomotor noise slope parameter for regressions\tabularnewline
$\mathrm{sre}_{1}^{\mathrm{(FS)}}$ & 5&0 & Saccadic range error intercept parameter for forward saccades\tabularnewline
$\mathrm{sre}_{2}^{\mathrm{(FS)}}$ & 0&5 & Saccadic range error slope parameter for forward saccades\tabularnewline
$\mathrm{sre}_{1}^{\mathrm{(SK)}}$ & 5&0 & Saccadic range error intercept parameter for skipping saccades\tabularnewline
$\mathrm{sre}_{2}^{\mathrm{(SK)}}$ & 0&75 & Saccadic range error slope parameter for skipping saccades\tabularnewline
$\mathrm{sre}_{1}^{\mathrm{(FRF)}}$ & 2&5 & Saccadic range error intercept parameter for forward refixations\tabularnewline
$\mathrm{sre}_{2}^{\mathrm{(FRF)}}$ & 0&5 & Saccadic range error slope parameter for forward refixations\tabularnewline
$\mathrm{sre}_{1}^{\mathrm{(BRF)}}$ & --2&5 & Saccadic range error intercept parameter for backward refixations\tabularnewline
$\mathrm{sre}_{2}^{\mathrm{(BRF)}}$ & --0&5 & Saccadic range error slope parameter for backward refixations\tabularnewline
$\mathrm{sre}_{1}^{\mathrm{(RG)}}$ & --2&5 & Saccadic range error intercept parameter for regressions\tabularnewline
$\mathrm{sre}_{2}^{\mathrm{(RG)}}$ & --0&9 & Saccadic range error slope parameter for regressions\tabularnewline
\bottomrule
\end{longtable}

\end{document}